\documentclass[final]{raa06}           
\usepackage{graphicx,times}             
\usepackage{natbib}
\usepackage{amssymb,amsmath}
\bibpunct{(}{)}{;}{a}{}{,}
\usepackage[colorlinks=true, citecolor=blue]{hyperref}%

\usepackage{graphicx,kantlipsum,setspace}
\usepackage{caption}
\usepackage{graphics,epsf}
\usepackage{amsmath}                
\usepackage{amsfonts}               
\usepackage{amssymb}                
\usepackage{epsfig}                 
\usepackage{appendix}
\usepackage{graphicx}
\usepackage{float}
\usepackage{color}
\usepackage{multirow}
\usepackage{colortbl}
\usepackage[para,online,flushleft]{threeparttable}
\usepackage{xcolor}

\hypersetup{citecolor=blue, 
            linkcolor=red, 
            menucolor=blue, 
            urlcolor=blue}  

\newcommand{\yr}{{~\rm yr}}

\begin{document}
\titlerunning{The jet-feedback mechanism in CEE}
   \title{The jet-feedback mechanism in common envelope evolution of planetary nebula progenitors
}

   \volnopage{Vol.0 (20xx) No.0, 000--000}      
   \setcounter{page}{1}          


   \author{Yonah Weiner, Noam Soker
    }

   \institute{Department of Physics, Technion - Israel Institute of Technology, Haifa, 3200003, Israel;   {\it    soker@physics.technion.ac.il} \\ 
   \vs\no
   {\small Noam Soker: orcid: {0000-0003-0375-8987}}\\ 
\vs\no
  }

\abstract{
Using the stellar evolution code \textsc{mesa}, we mimic the negative jet feedback mechanism in common envelope evolution (CEE) of low-mass main sequence stars, $M_2 \simeq 0.1-0.2 M_\odot$, spiraling inward inside the envelopes of asymptotic giant branch (AGB) or red giant branch (RGB) stars and find that the jets reduced the envelope density, therefore the jets' power, by a factor of $\chi \approx 0.5 (M_2/0.1 M_\odot)^{-1}$. We mimic the energy that the jets deposit into the envelope by depositing energy into the outer envelope, a process that inflates the envelope, therefore reducing the density in the vicinity of the main sequence star, the accretion rate, and the jets' power. In deriving this expression for the negative jet feedback coefficient $\chi$, we assume that the actual mass accretion rate is a fraction $\xi \approx 0.2-0.5$ of the classical Bondi-Hoyle-Lyttleton mass accretion rate and that the jets carry a fraction $\eta \approx 0.25-0.5$ of the accretion energy onto the main sequence star. Our study is another step in establishing the major role of jets in the onset and early phase of CEE,  a possible grazing envelope evolution phase, and in transient events, such as luminous red novae, which these processes can power.  
}

\maketitle

\section{Introduction}
\label{sec:Introduction}

Catalogs and large collections of planetary nebula (PN) images (e.g., \citealt{Balick1987, Chuetal1987, Schwarzetal1992, CorradiSchwarz1995, SahaiTrauger1998, Sahaietal2007, Sahaietal2011, Parkeretal2016, Parker2022}) show that many of them possess axisymmetrical, i.e., one symmetry axis, or multipolar, i.e., two or more symmetry axes, morphologies. On opposite sides along each symmetry axis, there might be pairs of dense clumps, ears, or lobes. Many studies consider pairs of jet to shape these pairs of structural features along symmetry axes (e.g., \citealt{Morris1987, Soker1990AJ, GarciaSegura1997, SahaiTrauger1998, GarciaSeguraLopez2000, GarciaSeguraetal2005, GarciaSeguraetal2021, GarciaSeguraetal2022, AkashiSoker2018, EstrellaTrujilloetal2019, Tafoyaetal2019, Balicketal2020, RechyGarciaetal2020, Clairmontetal2022, Danehkar2022, MoragaBaezetal2023, Derlopaetal2024, Mirandaetal2024, Sahaietal2024}); (\citealt{Baanetal2021} discuss an alternative shaping mechanism). We adopt the shaping by jet mechanism. 

We consider cases where a main-sequence companion accretes mass from an asymptotic giant branch (AGB) or, in rare cases ( e.g., \citealt{Hillwigetal2017, Sahaietal2017, Jonesetal2020, Jonesetal2022, Jonesetal2023}), a red giant branch (RGB) progenitor and launches opposite pairs of jets. A large fraction, likely the majority, of the binary progenitors of PNe with pairs of structural features along their symmetry axis or axes have evolved through a phase of common envelope evolution (CEE). In some the companion merged with the core or was tidally destroyed by the core; in others a close binary system (orbital periods of less than a few weeks) survived at the center of the PN, as evident from over a hundred post-AGB nebulae, pre-PNe, and PNe that have central binary systems (e.g., \citealt{Miszalski2019ic, Oroszetal2019, Jones2020Galax, Jones2025}). Based on the large fraction of jet-shaped PNe that have a post-CEE binary system at their center, \cite{Soker2025Robust} claimed that jets are the most robust observable ingredient of CEE. 

Some observations indicate that main-sequence stars can accrete mass during the CEE. The radii of some main-sequence stars in post-CEE binary systems at the center of PNe are larger than the radius of a main-sequence star of the same mass, suggesting mass accretion (e.g., \citealt{Jonesetal2015}).  \cite{BlackmanLucchini2014} argued that the large momenta of the nebulae of some PNe (e.g., \citealt{Bujarrabaletal2001, Sahaietal2008}) imply that the main-sequence companion launched jets during the CEE. Theoretically, mass accretion at a high rate without large stellar expansion is possible if the accretion onto the main-sequence star occurs via an accretion disk, and the jets launched from the disk remove the outer high-entropy layers of the newly accreted envelope (\citealt{BearSoker2025acc, Scolnicetal2025}). In that case, the main-sequence star accretes at a high rate while maintaining its deep potential well, ensuring relatively energetic jets. Observations show that the binary PN progenitor might launch jets before, during, or after the CEE (e.g.,  \citealt{Tocknelletal2014, Guerreroetal2020, Kimeswengeretal2021}). Theoretically, the binary system can launch jets in the grazing envelope evolution that might occur shortly before the CEE or might prevent the CEE (e.g., \citealt{Soker2020Galax}). 

The realization of the major role of jets in CEE motivated studies to simulate CEE with a companion launching jets inside the envelope, during a CEE or a grazing envelope evolution (e.g., \citealt{MorenoMendezetal2017, ShiberSoker2018, LopezCamaraetal2019, Shiberetal2019, LopezCamaraetal2020MN, Hilleletal2022, Hilleletal2023, LopezCamaraetal2022, Zouetal2022, Soker2022Rev, Schreieretal2023, Schreieretal2025, Gurjareta2024eas, ShiberIaconi2024}). Most of these studies, however, omit some processes or cover a short time of the CEE (less than the spiraling-in time). Despite the robust arguments for the crucial role of jets in CEE, the majority of CEE simulations do not include jets that the companion launches (e.g., \citealt{Staffetal2016MN, Kuruwitaetal2016, Ohlmannetal2016a,  Iaconietal2017b, Chamandyetal2019, LawSmithetal2020, GlanzPerets2021a, GlanzPerets2021b, GonzalezBolivaretal2022, GonzalezBolivaretal2024, Lauetal2022a, Lauetal2022b,  BermudezBustamanteetal2024, Chamandyetal2024, GagnierPejcha2024,  Landrietal2024, RosselliCalderon2024, Vetteretal2024, Bhattacharyyaetal2025, Vetteretal2025}, for some papers from the last decade). Despite considerable progress and efforts in recent years, the state of CEE research is still some distance from accurately simulating the CEE.

Generally, jets interact with the mass reservoir that feeds the accretion disk that launches the jets in a feedback mechanism. The feedback features both positive and negative components (see \citealt{Soker2016Rev} for a review). { 
In positive-feedback processes, the jets interact with the surroundings, allowing for higher mass accretion than if accretion were to take place without launching jets. For example, the accretion of angular momentum from a disk without angular momentum removal by jets will make the star spin too fast to maintain a hydrostatic equilibrium in its outer layers, hence stopping the accretion process. However, the removal of extra angular momentum by jets allows the accretion to continue. The jets carry a small fraction of the inflowing mass; hence, the mass of the accretor increases. Another outcome of accretion at a high rate is that a main-sequence star expands. This large expansion can halt accretion. The outer layers are of high entropy and energy. As the star expands, it covers the inner parts of the accretion disk that launch the jets. Under the assumption that the jets interact with the outer expanded layers and remove their mass, simulations with \textsc{mesa}, hence one-dimensional, that mimic this high-entropy mass removal demonstrated that the accretion can proceed while preventing large expansion of the main-sequence accretor (\citealt{BearSoker2025acc, Scolnicetal2025}). The jets and the mass they remove together are smaller than the inflowing mass, hence the accretor mass increases. Without the mass removal by jets, the expansion of the star halts accretion.   }
Overall, the positive feedback component is the jet-driven removal of angular momentum and energy from the immediate surroundings of the mass-accreting star, { but at a rate smaller than the mass inflow rate, } allowing more gas to flow in to continue the accretion process { at a higher rate than without the effects of the jets} (e.g., \citealt{Shiberetal2016, Chamandyetal2018}).

{ In the negative component of the feedback, the interaction of the jets with the surroundings reduces the accretion rate, hence the jets' power. }
Specifically, the energy and momentum that the jets deposit into the mass reservoir, which in the CEE is the giant's envelope, cause it to expand, thereby reducing the density and, consequently, the mass accretion rate. More powerful jets will reduce accretion more, and by regulating their power, this constitutes the negative feedback component (e.g., \citealt{Soker2016Rev}). 
{ In the scenario we consider in this study, the companion enters the giant's envelope with an accretion disk it builds while accreting mass via a Roche lobe overflow as it approaches the giant's envelope. We assume, like earlier studies, that the disk continues to survive in the giant's envelope, at least in the early CEE phases.  }
\cite{GrichenerCohenSoker2021} conduct one-dimensional simulations to quantitatively obtain the degree of the jet negative feedback in common envelope jet supernovae, which are CEE of a neutron star or a black hole inside the envelope of a red supergiant star, and later, possibly inside its core. \cite{Hilleletal2022} performed three-dimensional simulations, covering lower jet powers because of numerical limitations. 

In this study, we conduct one-dimensional simulations similar to those of \cite{GrichenerCohenSoker2021}, but for a main-sequence companion inside AGB and RGB stars. These systems are progenitors of some PNe. The goals are to reveal the general effect of the jets and, like \cite{GrichenerCohenSoker2021}, to set the stage for three-dimensional hydrodynamical simulations of the negative feedback mechanism. We describe the numerical method in Section \ref{sec:Mimicking}, our results in Section \ref{sec:Results}, and in Section \ref{sec:Neglecting} we discuss our assumption of neglecting the orbital energy of the binary system. In Section \ref{sec:Summary}, we summarize our study with a call to include jets in CEE simulations.  

\section{Mimicking the Negative Jet Feedback}
\label{sec:Mimicking}
As a companion main sequence star spirals in the common envelope of an AGB star, it accretes mass. We assume the companion launches jets, and we wish to estimate the feedback effect the jets have on the accretion rate. We do this by running a one-dimensional simulation of an AGB and an RGB stellar models. We list the assumptions and approximations below.

\subsection{Accretion Rate}
\label{subsec:AccretionRate}
 The mass accretion rate, $\dot M_{acc}$ , is a fraction $\xi$ of the Bondi-Hoyle-Lyttleton (BHL) mass accretion rate 
\begin{equation}
    \dot M_{\rm acc}= \xi \dot M_{\rm BHL}=\xi\pi R^2_{\rm acc}\rho v_{\rm rel} , 
 \label{eq:Macc1}   
\end{equation}
where $\dot M_{\rm BHL}$ is the BHL accretion rate, $R_{\rm acc}$ is the accretion radius, and $v_{\rm rel}$ is the velocity of the companion relative to the envelope. { The density $\rho$ in the classical BHL accretion flow is the unperturbed density at infinity; in the present study } we take the density to be the density at the AGB radius equals the orbital separation, i.e., the position of the companion.  We neglect the rotational velocity of the envelope so that $v_{\rm rel}\approx v_{\rm K}$ where $v_{\rm K}$ is the Keplerian velocity of the companion. We take the accretion radius to be 
\begin{equation}
    R_{\rm acc}=\frac{2GM_2}{v^2_{\rm rel}+C^2_{\rm s0}} \simeq \frac{2GM_2}{v_K ^2} \simeq \frac{2M_2a}{M(a)} ,
     \label{eq:Racc1}   
\end{equation}
where $C_{\rm s0}$ is the speed of sound of the unperturbed envelope, $a$ is the orbital radius of the companion, $M(a)$ is the mass of the primary star inner to the companion mass, and $M_2$ is the mass of the companion. We use these approximations. (1) We take $M_2 \ll M$, as in our simulations $M_2 = 0.2 M_\odot$ and $M=1.5M_\odot$. { The jet feedback mechanism works for more massive companions as well. However, since we are aiming to explore the pure role of jets, we neglect the orbital energy that the spiraling-in companion deposits in the envelope. This omission is justified only for low-mass companions, as we elaborate on later. } (2) We neglect the sound speed that otherwise reduces mass accretion rate, as we also neglect envelope rotation, which would increase accretion rate by reducing $v_{\rm rel}$. We assume these two effects canceled out for the accuracy of our simulations. (3) We take $M(a)\approx M$ because our simulation takes place in the envelope, and a large fraction of the primary mass is in its core. These assumptions introduce some uncertainties, which are absorbed in the efficiency parameter $\xi$ (equation \ref{eq:Macc1}).

Three-dimensional (3D) hydrodynamic simulations by \cite{Chamandyetal2018} have shown that if the companion launches jets, bipolar structures of low-density gas are inflated, forming a `cocoon'. The effect of the jets is to reduce the envelope density by a factor of  $\chi=\rho/\rho_0 < 1$. This, in turn, reduces the mass accretion rate and the jet power, i.e., a negative feedback cycle. The new accretion rate is 
\begin{equation}
     \dot M_{\rm acc}=\chi \dot M_{\rm acc,0} =\chi \xi M_{\rm BHL,0}.
\label{eq:Macc2}
\end{equation}
The "$0$" subscript refers to the unperturbed envelope, i.e., taking $\rho_0$ in equation (\ref{eq:Macc1}). Calculating $\chi$ via 1D simulations is the objective of this study.

\subsection{The Power of the Jets}
\label{subsec:Power}

Because the envelope is optically thick and radiation cannot escape freely, we assume that the jets take a large fraction, $\eta$,  of the accretion energy. We estimate the energy of the jets from the virial theorem to be
\begin{equation}
\begin{split}
   \dot E_{\rm 2j} & = \eta \frac{G M_2 \dot M_{\rm acc}}{2R_{2}}=\eta\xi \chi \frac{G M_2 \dot M_{\rm BHL,0}}{2R_{2}} 
   \\ & = \zeta \frac{G M _2\dot M_{\rm BHL,0}}{2R_{2}}, 
\label{eq:E2j}
\end{split}
\end{equation}
where in the last equality we defined
\begin{equation}
    \zeta\equiv\eta\xi\chi
    \label{eq:zeta},
\end{equation} 
and $R_2$ is the radius of the mass-accreting companion. 
By expanding equation (\ref{eq:E2j}) under our assumptions and approximations, and taking  $v_{\rm rel} \simeq \sqrt{G(M+M_2)/a}$ in equation (\ref{eq:Macc1}),  we have 
\begin{equation} 
    \dot E_{2j} =2\pi \zeta G  \frac{M_2^3a^2}{M^2R_{2}}   \rho_0 \sqrt{\frac{G (M+M_2) }{a}}.
    \label{eq:E2jFull}
\end{equation}
This is the power of the two { opposite jets that the companion launches } that we deposit into the stellar envelope model of our simulations. 
We do not include the orbital energy that the spiraling-in companion deposits. We discuss this assumption in Section \ref{sec:Neglecting}. 

\subsection{The spiraling-in timescale}
\label{subsec:TimeScale}

Hydrodynamical simulations of the CEE (section \ref{sec:Introduction}), show that the time it takes for the companion to spiral in is of the order of a Keplerian period $P$ on the surface. For our simulations, we take a constant inward radial velocity (defined to be positive) of the in-spiraling companion (see also \citealt{GrichenerCohenSoker2021}) 
\begin{equation}
    v_{\rm in}=\frac{R}{2P},
    \label{radial velocity}
\end{equation}
where $R$ is the radius of the giant star.  For the AGB star $v_{\rm in}=0.3R_\odot ~ {\rm day}^{-1} =110 R_\odot \yr^{-1}$ and for the RGB star $v_{\rm in}=0.5R_\odot ~ {\rm day}^{-1} =183 R_\odot \yr^{-1}$. 

\subsection{The Simulations}
\label{subsec:Simulations}

We use the one-dimensional stellar evolution code \textsc{mesa} (Modules for Experiments in Stellar Astrophysics; version r24.08; \citealt{Paxton2011, Paxton2013, Paxton2015, Paxton2018, Paxton2019, Jermyn2023}), and deposit energy according to equation (\ref{eq:E2jFull}) to explore the jet feedback mechanism. 
We simulate the jet feedback mechanism in AGB and RGB stellar models. 

\subsubsection{The AGB Simulations}
\label{subsubsec:AGBSimulation}
We built an AGB stellar model from a zero-age main-sequence (ZAMS) star of  $M_{\rm ZAMS}=1.5M_\odot$  and a metallicity of $Z=0.02$. We assume that the CEE begins during a helium-shell flash (thermal pulse) on the AGB, which causes envelope expansion. We, therefore, use the AGB stellar model during one of the helium-shell flashes, when the radius is at its maximum, which was $R=308 R_\odot$. The star also reached a mass of $M=1.18M_\odot$ due to stellar winds. At this point, we assume a low-mass main-sequence companion with mass of $M_2=0.2M_\odot$ and radius $R_2=0.23R_\odot$ enters the envelope of the AGB star.

To simulate the energy the jets deposit into the envelope, we continue to evolve the star while adding energy to the envelope according to equation (\ref{eq:E2jFull}) for a given value of $\zeta$. At every timestep $\Delta t$, we add an energy of $\Delta E = \dot E_{\rm 2j} \Delta t$ into a thick envelope shell of mass $M_{\rm TS}$. To each numerical shell $\Delta m_{i}$, we add an energy of $\Delta E_{i}=(\Delta M_{i}/M_{TS})\Delta E$. At the beginning of the in-spiral { when the orbital separation is $a>200 R_\odot$, } the thick shell of mass $M_{\rm TS}$ { into which we deposit the jets' energy, } is bounded by the photosphere and the inner radius of $200R_\odot$. We use this inner radius, rather than an outer one { near the location of the companion close to the photosphere, } to prevent numerical difficulties that would arise if we injected the energy into a small envelope mass. Additionally, the jets spread the energy over a larger volume. In any case, we consider our results less accurate when the companion is in the very outer zones of the envelope. After the companion reaches $a=200R_\odot$, { which is the inner boundary of the shell into which we deposit energy at early CEE phases, } the inner boundary of the shell is located at the companion star at $a(t)=R-v_{\rm in}t < 200 R_\odot$. 

We terminate the simulation when the companion reaches $20\%$ of the stellar radius, because we expect other effects to become important, like the orbital energy we neglect. We consider the stellar density profile at this time, $\rho(r)$, as the perturbed envelope density, and use it to calculate the parameter $\chi=\rho/\rho_0$. We present the value of $\chi$ at three radii, { $100$, $200$, and $300 R_\odot$. }
We repeat simulations for different values of $\zeta$, in the range of $\zeta=0.002-0.016$ with intervals of $\simeq 0.002$; the maximum value is set by numerical limitations (Section \ref{subsec:limitations}). The output of the simulations is $\chi(r,\zeta)$. 

\subsubsection{The RGB Simulations}
\label{subsubsec:RGBSimulation}

We also ran simulations for jets inside of an RGB envelope. We built an $M=1.5M_\odot$ ZAMS star exactly like in the AGB case. We evolved the stellar model until its RGB stage when its radius reached $R=163R_\odot$; due to mass loss, its mass is  $M=1.38M_\odot$. We assume at this point that the companion star enters the envelope of the RGB star. We then continue the evolution while depositing the jets' energy, according to equation (\ref{eq:E2jFull}) as in the AGB case. In the RGB simulations, the inner boundary of the jet energy injection shell was initially at $R=100R_\odot$. After the companion spirals further in than $R=100R_\odot$, the shell inner boundary is at $a(t)=R-v_{\rm in} t$.

We repeated simulations for different $\zeta$ values, in the range of $\zeta=0.002-0.019$  with intervals of $\zeta\simeq 0.002$ . The output of the simulations is $\chi(r,\zeta)$. 

\subsection{Numerical Limitations}
\label{subsec:limitations}

Ideally, we wanted to simulate higher values of $\zeta$ of around $\zeta \simeq 0.1$. We found, however, that the \textsc{mesa} code failed to converge (hence, stop running) for $\zeta > 0.016$ in the AGB simulations and $\zeta > 0.019 $ in the RGB simulations.  

Injecting energy expands the envelope, sometimes outer zones reached the escape velocity or higher; the hydrostatic equilibrium assumption breaks down. The \textsc{mesa} code can cope with this problem by choosing the hydrodynamic mode instead of the hydrostatic one.  This change allowed us to reach a higher value of $\zeta$ as mentioned above.

\section{The negative jet feedback}
\label{sec:Results}

\begin{figure}[]
	\begin{center}
\includegraphics[trim=1.0cm 0.0cm 0.0cm 1.0cm ,clip, scale=0.28]{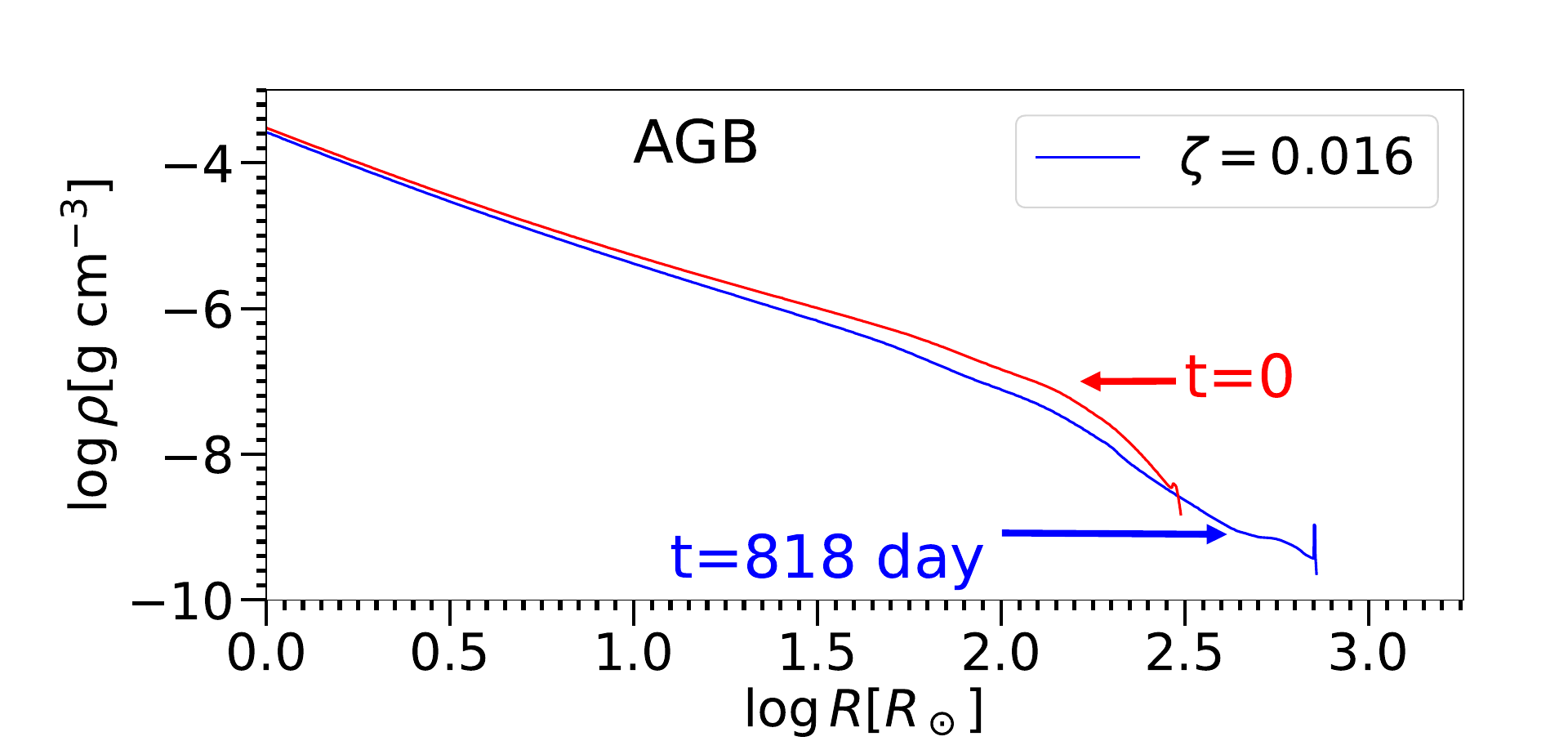} 
\caption{The AGB stellar density profiles at $t=0$ when we start energy injection (red line), and at $t=818~{\rm day}$ (blue line), in a model with a companion star of $M_2=0.2 M_\odot$ and $\zeta=0.016$ in equation (\ref{eq:E2jFull}). 
As expected, the star expands; in this case by a factor of of $2.3$. Note the density inversion very close to the photosphere. }
\label{Fig:AGB Density}
\end{center}
\end{figure}

\begin{figure}[]
	\begin{center}
\includegraphics[trim=1.0cm 0.0cm 0.0cm 1.0cm ,clip, scale=0.28]{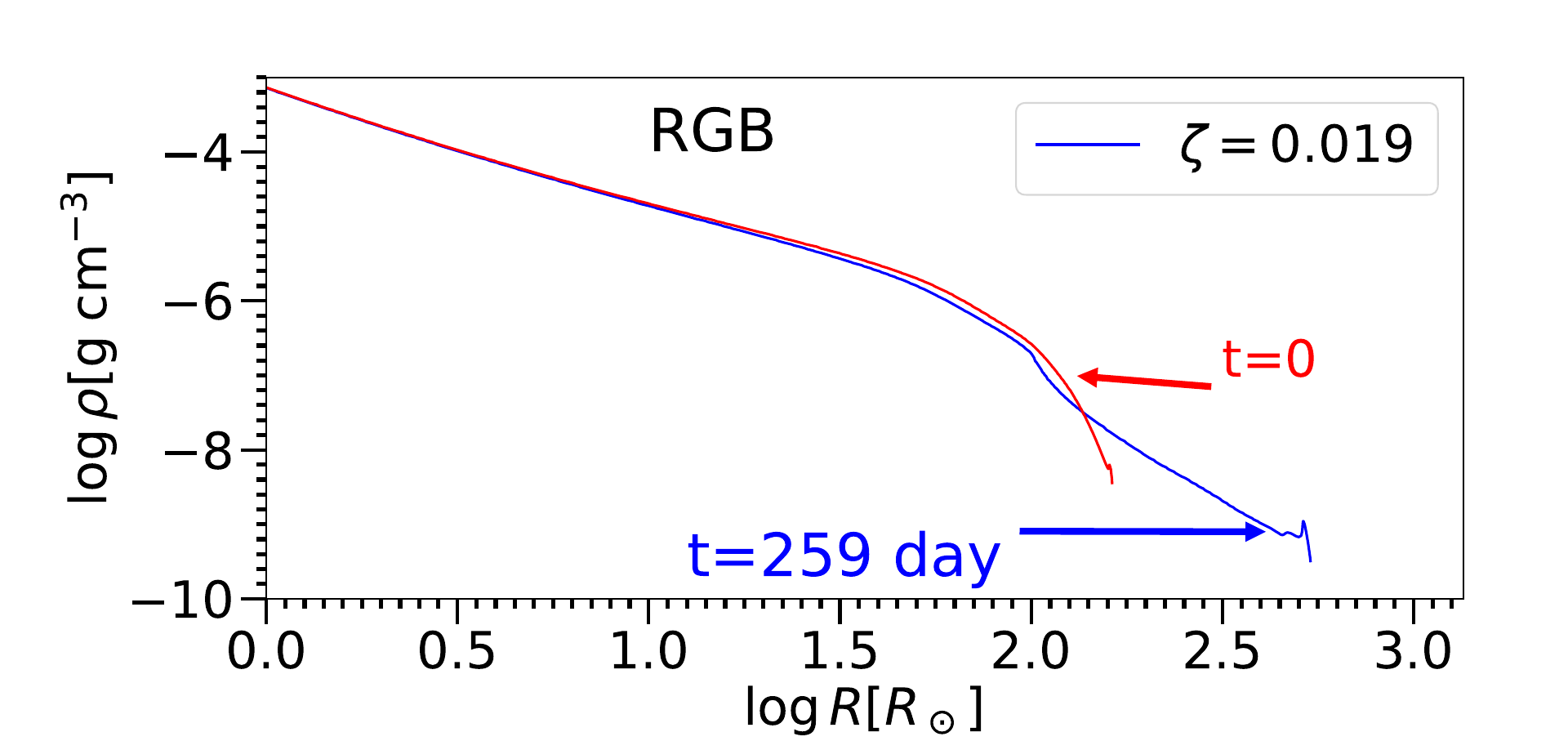} 
\caption{The RGB stellar density profiles at $t=0$ when we start energy injection (red line), and at $t=259~{\rm day}$, in a model with a companion star of $M_2=0.2 M_\odot$ and $\zeta=0.019$ in equation (\ref{eq:E2jFull}). 
The star expands by a factor of $3.3$ for these parameters and this time. 
}
\label{Fig:RGB Density}
\end{center}
\end{figure}

As we inject energy, the envelope expands { and the density at and near the orbit of the companion decreases. We present these effects for a one-dimensional simulation of an AGB star } in a CEE with a companion of mass $M_2=0.2 M_\odot$ and taking $\zeta=0.016$ in equation (\ref{eq:E2jFull}). In Figure \ref{Fig:AGB Density} we present the density profiles just before energy injection, $t=0$, and at $t=818~{\rm day}$. The AGB star expands by a factor of $2.3$ between these two times. 
{ For these parameters, $\zeta=0.016$ and $t=818~{\rm day}$, the main-sequence companion accreted a mass of $M_{\rm acc} = 0.009 (\eta/0.5)^{-1} M_\odot$. This is small, but non-negligible, compared to the companion initial mass, so that we expect the main-sequence companion to expand somewhat.  }
In Figure \ref{Fig:RGB Density} we present two profiles for an RGB model with the same companion, but in this case we could simulate a value of $\zeta=0.19$; the profiles are at $t=0$ and  $t=259~{\rm day}$. Again, as expected, the RGB envelope expands by a factor of $3.3$. 
{ For these parameters, $\zeta=0.019$ and $t=259~{\rm day}$, the main-sequence companion accreted a mass of $M_{\rm acc} = 0.012 (\eta/0.5)^{-1} M_\odot$. Again, this is a small mass, but non-negligible, compared to the companion's initial mass, so that we expect it to expand somewhat.  }

\begin{figure*}[]
	\begin{center}
\includegraphics[trim=1.0cm 0.0cm 0.0cm 1.0cm ,clip, scale=0.60]{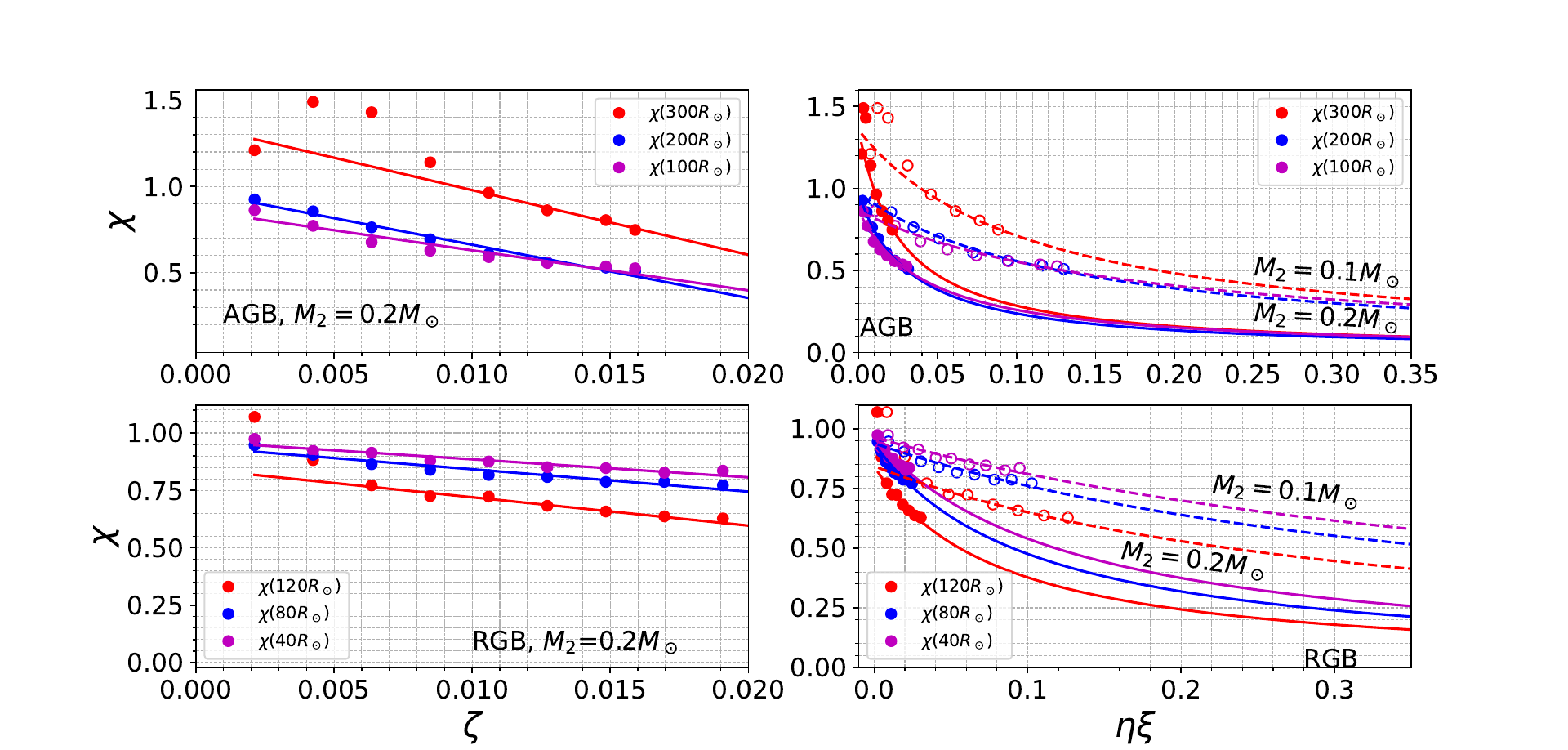} 
\caption{Left panels: The density change factor $\chi$ at three radii { (given in the insets) } as a function of $\zeta$ for a companion of $M_2=0.2 M_\odot$, and for an AGB stellar model at $t=818~{\rm day}$ (upper panels) and an RGB stellar model at $t=259~{\rm day}$ (lower panels); insets give the radii and color code. The solid lines, with color corresponding to the points, are the best linear fit of $\chi(\zeta)$.  The filled circles in the right panels present the same simulations and for $M_2=0.2 M_\odot$ as in the corresponding panels on the left, but in the plane of $\chi$ versus $\eta \xi = \zeta/\chi$. The solid lines are the best fits from the left panel, extrapolated to higher value of $\eta \xi$ than we simulated. The open circles are the same simulations, but $\zeta$ and $\eta \xi$ are for $M_1=0.1M_\odot$ by equation (\ref{eq:E2jFull}). The dotted lines are the best fits and extrapolation to the case of $M_1=0.1M_\odot$. }
\label{Fig:linear fit}
\end{center}
\end{figure*}

To explore the negative jet feedback, we conduct several simulations with different values of $\zeta$, up to the maximum value we could simulate; for higher values of $\zeta$ for a given $M_2$, the code does not converge. The numerical limit is the power of the energy we inject. Therefore, by equation (\ref{eq:E2jFull}), for lower secondary masses $M_2$, the maximum possible $\zeta$ value is larger.  
The set of simulations gives us the decrease (in the relevant simulations) in the envelope density, $\chi$, as a function of $\zeta$, radius, and time,   $\chi(\zeta, r, t)$. For each simulation with a prescribed value of $\zeta$, we record $\chi(\zeta, r, t)$ at the end of the simulation when the companion reached an orbital separation of $a=0.2R_0$, where $R_0$ is the radius of the AGB or RGB envelope at $t=0$, and at three radii. We present these for CEE of an AGB and an RGB stellar models with an $M_2=0.2 M_\odot$ companion in the left panels of Figure \ref{Fig:linear fit}. The insets give the three radii (different for the AGB and RGB simulations). For each radius, we make a linear fit of $\chi(\zeta)$. { We did not include points where $\chi>1$ in the fitting (relevant to the outer radius in each stellar model), } as by the end of the simulation, the companion is deep inside the envelope and the outer radius is not relevant for the accretion process. 

Since we cannot simulate higher values of $\zeta$ we extrapolate our results. In the right panels of Figure \ref{Fig:linear fit}, we plot $\chi$ as a function of $\eta\xi = \zeta/\chi$, where $\eta$ is the fraction of the accretion energy that the jets carry (equation \ref{eq:E2j}), and $\xi$ is the accretion rate relative to the classical BHL rate (equation \ref{eq:Macc1}). We continue the lines (extrapolate) to higher values of $\eta \xi$ by the linear fit of the panels on the left; these zones correspond to higher values of $\zeta$ than we could simulate. We assume that $\eta \approx 0.25-0.5$ as some accretion energy goes to inflate the companion. We take $\xi \approx 0.2-0.5$ based on the high-quality three-dimensional simulations by \cite{Kashietal2022}, who find $\xi \simeq 0.5$ (also, e.g., \citealt{Livio1986}). We are aware that some simulations yield significantly lower values (e.g., \citealt{Ricker2008, MacLeodRamirezRuiz2015, MacLeodetal2017}), { from $\simeq 0.5$ times the BHL rate for a uniform density, down to $\simeq 0.01$ times the BHL accretion rate for a steep density gradient perpendicular to the relative velocity of the star and the gas (e.g., \citealt{MacLeodRamirezRuiz2015}). } Therefore, we allow a somewhat lower value than the finding of \cite{Kashietal2022}. 
We are interested, therefore, in the range of $\eta \xi \approx 0.05-0.25$. 

The solid lines in the right panels of Figure \ref{Fig:linear fit} are the fitting of the left panels with their extrapolation. For $M_2=0.2 M_\odot$ and the range of $\eta \xi \approx 0.05-0.25$, we find the negative jet feedback coefficient to be $\chi_{\rm AGB,0.2} \simeq 0.1-0.4$ and $\chi_{\rm RGB,0.2} \simeq 0.3-0.7$ (we give a little weight to the outer radius represented by the red lines).  

In the way we define the negative jet feedback coefficient, and taking $M_2/R_2$ to be about constant for different low-mass main sequence stars, equation (\ref{eq:E2jFull}) shows that approximately $\zeta \propto M^{-2}_2$, hence for a given value of $\chi$, the approximate relation $\eta \xi \propto M^{-2}_2$ holds.  
The dashed lines in the right panels of Figure \ref{Fig:linear fit} are for $M_2=0.1 M_\odot$, and therefore the points from the simulations extend to values of $\eta \xi$ that are about four times larger than for $M_2=0.2 M_\odot$. We recall that in the simulations themselves, we inject energy. We translate the energy injection rate to a value of $\zeta$ by choosing $M_2$, calculating its radius $R_2$, and substituting in equation (\ref{eq:E2jFull}). We can vary the value of $M_2$. For $M_2=0.1M_\odot$ we find from the right panels of Figure \ref{Fig:linear fit}, and for the range of $\eta \xi \approx 0.05-0.25$, $\chi_{\rm AGB,0.1} \simeq 0.4-0.7$ and $\chi_{\rm RGB,0.1} \simeq 0.6-0.9$. 

To the accuracy of the present simulations, and at the middle of the $\eta \xi$ range, i.e., $\eta \xi \simeq 0.15$, we find the negative jet feedback coefficient for low-mass main sequence stars using the linear fit to crudely be 
\begin{equation}
\chi_{\rm AGB} \simeq 0.5 \left( \frac{M_2}{0.1M_\odot} \right)^{-1}, 
    \label{eq:FitAGB}
\end{equation}
and
\begin{equation}
\chi_{\rm RGB} \simeq 0.8 \left( \frac{M_2}{0.1M_\odot} \right)^{-1}  .
    \label{eq:FitRGB}
\end{equation}

\begin{figure*}[]
	\begin{center}
\includegraphics[trim=1.0cm 0.0cm 0.0cm 1.0cm ,clip, scale=0.60]{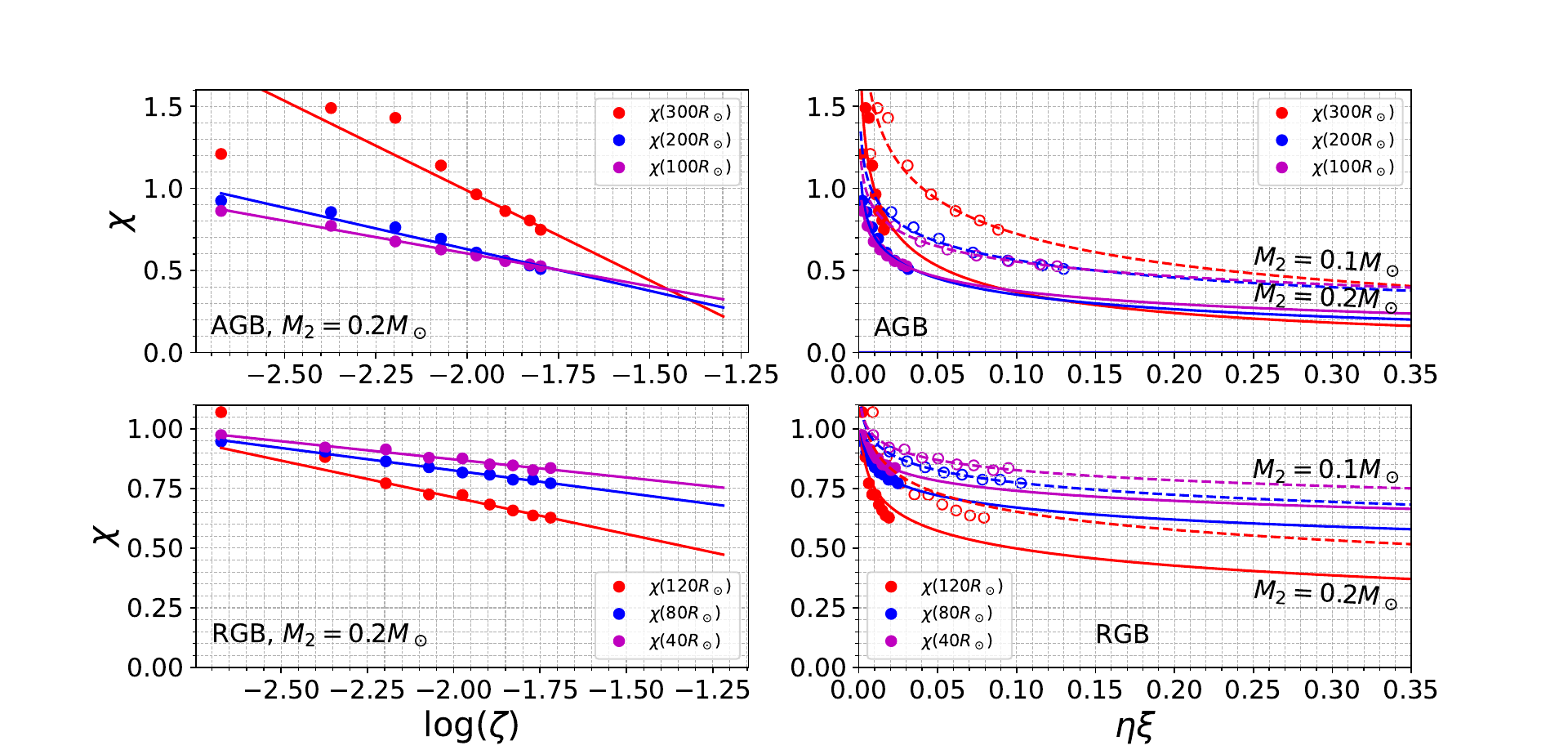} 
\caption{Similar to Figure \ref{Fig:linear fit} and for the same simulations, but in the left panels we fit with a linear line for $\chi (\log \zeta$). We use the extrapolations of these log fittings in the right panels.   }
\label{Fig:log fit}
\end{center}
\end{figure*}

Since we extrapolate to values of $\zeta$ that are much larger than what we can simulate, implying large uncertainties, we also extrapolate { our results } by performing a linear fit to $\chi$ as a function of $\log \zeta$ (the log fit). We present the log fittings in the left panels of Figure \ref{Fig:log fit} and their extrapolations in the right panels of Figure \ref{Fig:log fit}. From the log fittings, and with the same assumptions for $\eta$ and $\xi$ as above, we conclude from Figure \ref{Fig:log fit} that for $M_2=0.2 M_\odot$, $\chi_{AGB,0.2}=0.2-0.5$ and $\chi_{RGB,0.2}=0.6-0.8$ (we give little weight to the outer radius represented by the red lines). 

With the values of $\chi$ for $M_2=0.2M_\odot$ and $M_2=0.1M_\odot$, we find near $\eta \xi \simeq 0.15$, $\chi_{\rm AGB} \simeq 0.5 (M_2/0.1M_\odot)^{-1}$, similar to the result with the linear fit. For the RGB, we ignore the fitting to the outer radius (red lines) that, in any case, are far from the secondary star at the end of the simulation, and find  $\chi_{\rm RGB} \approx 0.7$. 

Our crude estimates warrant further studies, particularly three-dimensional simulations. Nonetheless, we propose using the crude expressions (\ref{eq:FitAGB}) and (\ref{eq:FitRGB}) in studies that require values of $\chi$, such as population synthesis studies of the jet feedback mechanism. 

\section{On the omission of the orbital Energy}
\label{sec:Neglecting}

As the companion spirals in, the binary system loses gravitational energy to the common envelope at a rate of $\dot E_{\rm B}$. In the present study, we did not consider this energy because our aim was to investigate the sole role of accretion energy. 
For $\dot E_{\rm B}$ we use the expression
\begin{equation}
\begin{split}
     \dot E_{\rm B} &= \frac{dE_{\rm B}}{da}\frac{da}{dt}=\frac{GM(a)M_2}{2a^2}v_{\rm in}=\frac{GM(a)M_2}{2a^2}\frac{R}{2P}\\
     &=\frac{1}{8\pi}\frac{GM(a)M_2R^2}{a^2} \sqrt{\frac{R}{G(M+M_2)}},
\end{split}
\label{eq:Eb}
\end{equation}
where $a$ is the orbital separation, and $v_{\rm in}$ is the inward radial velocity from equation (\ref{radial velocity}).

\begin{figure}[]
	\begin{center}

\includegraphics[trim=0.5cm 0.0cm 0.0cm 1.0cm ,clip, scale=0.28]{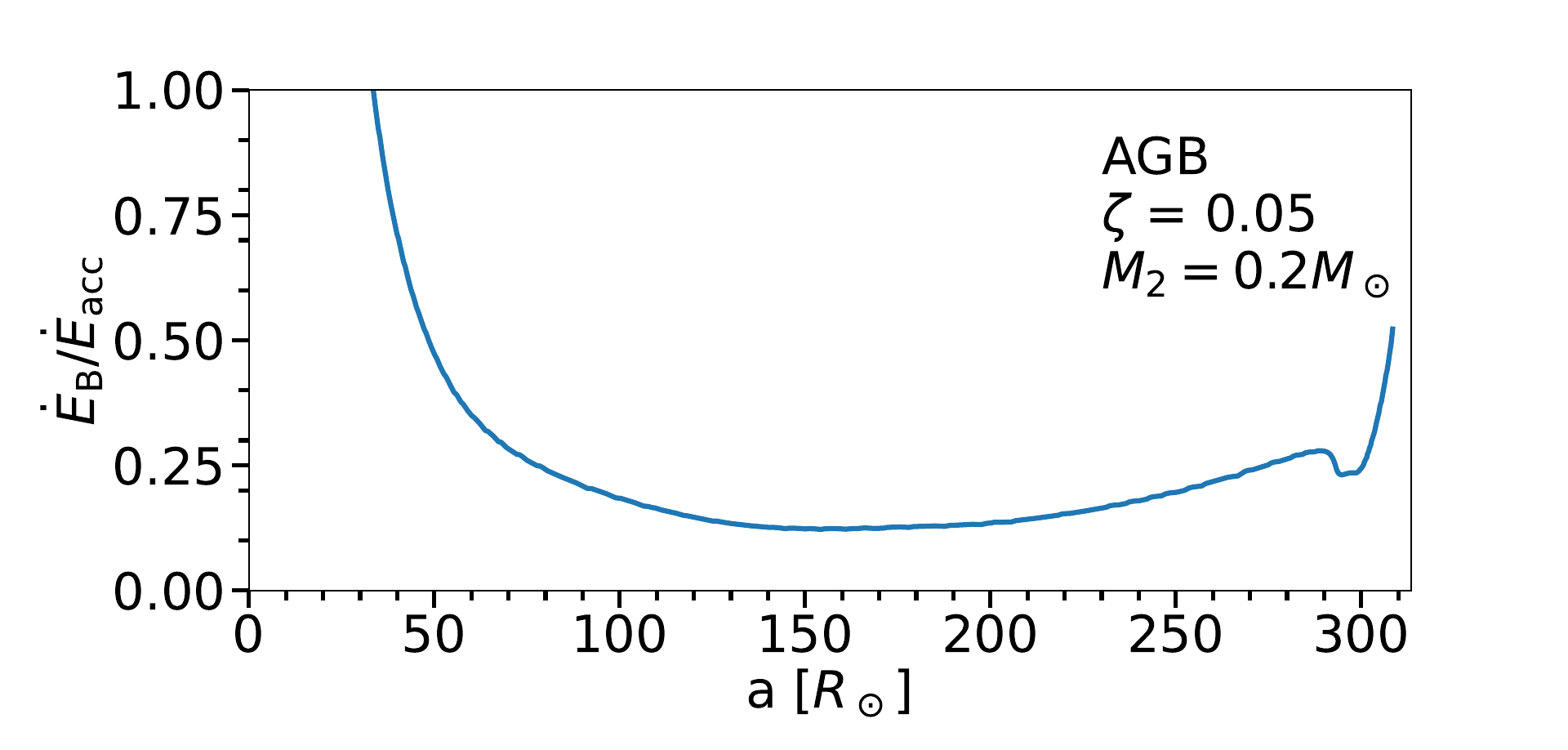} 
\caption{The ratio of the power of the gravitational energy that the binary system releases according to equation (\ref{eq:Eb}), to the jets' power, for a simulation with $M_2=0.2 M_\odot$ and $\zeta=0.05$. In regions where this ratio is much lower than unity, our assumption of neglecting the gravitational energy of the binary system is justified.  }
\label{Fig:gravjetAGB}
\end{center}
\end{figure}

\begin{figure}[]
	\begin{center}

\includegraphics[trim=1.0cm 0.0cm 0.0cm 1.0cm ,clip, scale=0.28]{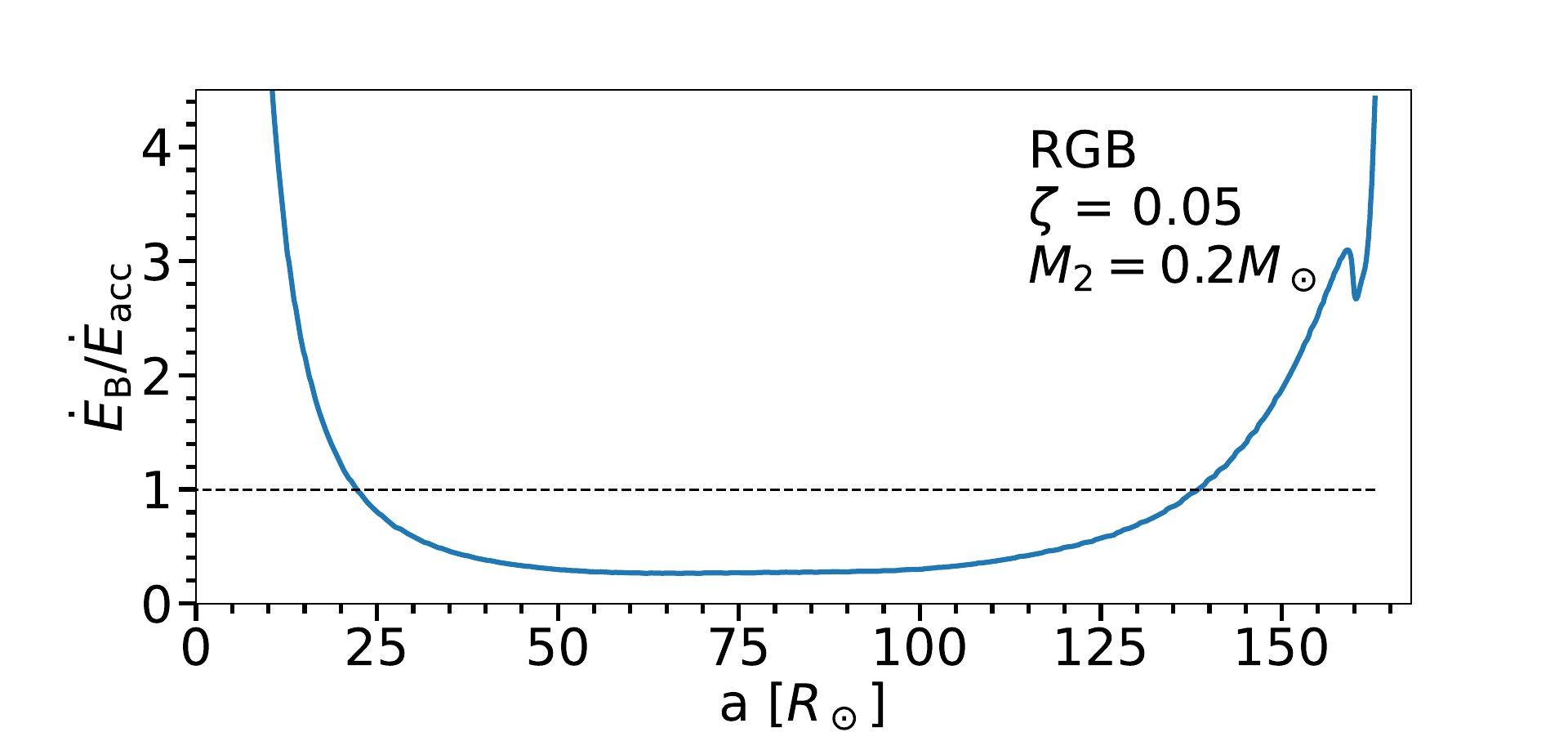} 
\caption{Similar to Figure \ref{Fig:gravjetAGB}, but for the RGB stellar model. { The dotted horizontal line marks equal power of orbital energy and jets. } }
\label{Fig:gravjetRGB}
\end{center}
\end{figure}

We plotted in Figures \ref{Fig:gravjetAGB} and \ref{Fig:gravjetRGB} the ratio $\dot E_{\rm B}/\dot E_{\rm acc}$ as a function of orbital separation $a$, and for $\zeta=0.05$ with $M_2=0.2 M_\odot$. The ratio $\dot E_{\rm B}/\dot E_{\rm acc}$ in the very outer envelope is not correct because we only consider here BHL accretion.
{ The BHL accretion assumes that the gas surrounds the accreting mass from all directions, at least to a distance of the accretion radius from all directions. When the companion is in the very outer envelope, the distance of the companion from the stellar radius is smaller than the accretion radius; the assumptions of the BHL accretion do not hold in full. }
However, when the companion enters the envelope, it accretes at a high rate via a Roche lobe overflow.  

In the case of the AGB stellar model we use, and for a low mass main sequence companion with the expected value of $\zeta$, Figure \ref{Fig:gravjetAGB} shows that we could neglect the spiraling-in energy until the companion reaches down to an orbital separation of $a \lesssim 40R_\odot$.
This is compatible with the view that in cases of main-sequence and white dwarfs companions, the jets play a major role just before the CEE, possibly in a grazing envelope evolution, and during the early CEE, i.e., when the companion is in the outer zones of the giant's envelope (e.g., \citealt{Soker2025WDfeed}). 
Figure \ref{Fig:gravjetRGB} shows that in the RGB case, our omission of the binary gravitational energy is justified down to a radius of $a \lesssim 25 R_\odot$. 

\section{Summary}
\label{sec:Summary}

We conducted one-dimensional simulations to estimate the degree to which jets reduce the density in the envelope, thereby affecting the mass accretion rate and their power, in a CEE of a low-mass main-sequence star in the envelope of an AGB or an RGB star. We used the \textsc{mesa} code to inject energy, mimicking the jets, into the envelope and examine the factor $\chi$ by which the envelope density decreases. The spherically symmetric code we use is limited in simulating the effect of jets that a companion launches as it orbits within the envelope, because the jets-envelope interaction forms a highly non-spherical flow. In addition, due to numerical limitations, we were unable to inject energy at the high power expected in the CEE according to the jet feedback mechanism. We had to extrapolate, as shown in the right panels of Figures \ref{Fig:linear fit} and \ref{Fig:log fit}. Taking the actual mass accretion rate to be lower than the classical BHL accretion rate by a factor $\xi \approx 0.2-0.5$ and the fraction of accretion energy that the jets carry to be $\eta \approx 0.25-0.5$, we examine the range $\eta \xi \approx 0.05-0.25$ in the extrapolated zones (the justifications are in Section \ref{sec:Results}). 
Given these limitations and uncertainties, we do not expect our results to be entirely accurate. Nonetheless, our results crudely demonstrate the negative jet feedback effect.  

The jets reduce the envelope density in the companion's vicinity and, consequently, the jets' power by a factor summarized in equations (\ref{eq:FitAGB}) and (\ref{eq:FitRGB}) for AGB and RGB stars, respectively. 
These two equations are our primary result. 

The next step would be to improve expressions (\ref{eq:FitAGB}) and (\ref{eq:FitRGB}) by adding the gravitational energy of the binary system. This is beyond the scope of this study, as such simulations should be conducted in three dimensions, as \cite{Hilleletal2022} did for a neutron star inside the envelope of a red supergiant. 
The jets also have a positive feedback component, removing energy and angular momentum from the close vicinity of the mass-accreting star, which allows for high mass-accretion rates. We did not calculate this effect in this study. Future CEE simulations should also include this effect. 

\cite{GrichenerCohenSoker2021} who simulated with \textsc{mesa} the jet feedback mechanism of a neutron star inside the envelope of a red supergiant star, i.e., a common envelope jet supernovae event, found $\chi_{\rm NS} \simeq 0.04-0.3$. \cite{Hilleletal2022} simulate this process with a three-dimensional hydrodynamical code in a regime of lower values of $\zeta$. They could fit a line through the two regimes, despite the different numerical code and settings. This fit reads $\chi_{\rm NS} = -0.2 \log \zeta - 0.34$. 
Since the accretion process onto a neutron star at high rates cools by neutrino emission, the fraction of accretion energy carried by the jets is much lower, and they assume $ \eta \xi \simeq 0.01-0.05$. From these, \cite{Hilleletal2022} found $\chi_{\rm NS} \simeq 0.1-0.2$. We conclude that jets that a neutron star launches have larger effects in reducing mass accretion rate in the negative jet feedback mechanism than typical expected jets that a main sequence companion launches.  

Despite the strong assumptions and some omissions of this study, including the assumption of spherical symmetry and hydrostatic equilibrium (although the CEE is not), due to the limitations of the 1D code \textsc{mesa}, and the omission of binary gravitational energy (which we justified), this study has merit for the following reasons. (1) We use the same approach as \cite{GrichenerCohenSoker2021}, with the same assumptions and omissions, but for a main sequence rather than an NS companion. \cite{Hilleletal2022} performed 3D simulations that included non-spherical and hydrodynamical flow with an NS companion. The results of these two studies cover different ranges of $\xi$, but are on the same linear fit! This indicates that the 1D \text{mesa} simulations, despite their assumptions and omissions, yield reasonable results. We intend to conduct 3D hydrodynamical simulations with a main-sequence companion rather than an NS. This paper, though, must be the first step. (2) The CEE is a critical phase in the evolution of many systems. The CEE attracts a huge amount of attention, like in the formation of close compact binary systems (e.g., \citealt{Xueetal2025RAA, Zhengetal2025RAA}) and different kinds of white dwarf binaries (e.g., \citealt{Zhouetal2015RAA, Zhuetal2023, Qietal2023}), so that even our simplified study of the jet feedback mechanism in CEE can have an impact relevant to many astrophysical systems. One of our goals is to promote more simulations of the jet feedback mechanism in CEE.  

We end by strengthening the call to consider jets that the more compact companion, the mass-accreting star, 
launches as it enters a CEE and at the early CEE phases (see also \citealt{Soker2023RAA}). Just before the CEE, the system might experience the grazing envelope evolution phase, where jets efficiently remove the envelope outskirts. These early phases might power a transient event, particularly luminous red novae. Jets might be necessary to power energetic luminous red novae.

\section*{Acknowledgements}

{We thank an anonymous referee for suggestions and comments that clarified and improved the paper.}
A grant from the Pazy Foundation supported this research. N.S. thanks the Charles Wolfson Academic Chair at the Technion for the support.


\begin{thebibliography}{}
\makeatletter
\relax
\def\mn@urlcharsother{\let\do\@makeother \do\$\do\&\do\#\do\^\do\_\do\%\do\~}
\def\mn@doi{\begingroup\mn@urlcharsother \@ifnextchar [ {\mn@doi@} {\mn@doi@[]}}
\def\mn@doi@[#1]#2{\def\@tempa{#1}\ifx\@tempa\@empty \href {http://dx.doi.org/#2} {doi:#2}\else \href {http://dx.doi.org/#2} {#1}\fi \endgroup}
\def\mn@eprint#1#2{\mn@eprint@#1:#2::\@nil}
\def\mn@eprint@arXiv#1{\href {http://arxiv.org/abs/#1} {{\tt arXiv:#1}}}
\def\mn@eprint@dblp#1{\href {http://dblp.uni-trier.de/rec/bibtex/#1.xml} {dblp:#1}}
\def\mn@eprint@#1:#2:#3:#4\@nil{\def\@tempa {#1}\def\@tempb {#2}\def\@tempc {#3}\ifx \@tempc \@empty \let \@tempc \@tempb \let \@tempb \@tempa \fi \ifx \@tempb \@empty \def\@tempb {arXiv}\fi \@ifundefined {mn@eprint@\@tempb}{\@tempb:\@tempc}{\expandafter \expandafter \csname mn@eprint@\@tempb\endcsname \expandafter{\@tempc}}}

\bibitem[\protect\citeauthoryear{{Akashi} \& {Soker}}{{Akashi} \& {Soker}}{2018}]{AkashiSoker2018}
{Akashi} M.,  {Soker} N.,  2018, \mn@doi [\mnras] {10.1093/mnras/sty2479}, \href {https://ui.adsabs.harvard.edu/abs/2018MNRAS.481.2754A} {481, 2754}

\bibitem[\protect\citeauthoryear{{Baan}, {Imai}  \& {Orosz}}{{Baan} et~al.}{2021}]{Baanetal2021}
{Baan} W.~A.,  {Imai} H.,   {Orosz} G.,  2021, \mn@doi [Research in Astronomy and Astrophysics] {10.1088/1674-4527/21/11/275}, \href {https://ui.adsabs.harvard.edu/abs/2021RAA....21..275B} {21, 275}

\bibitem[\protect\citeauthoryear{{Balick}}{{Balick}}{1987}]{Balick1987}
{Balick} B.,  1987, \mn@doi [\aj] {10.1086/114504}, \href {https://ui.adsabs.harvard.edu/abs/1987AJ.....94..671B} {94, 671}

\bibitem[\protect\citeauthoryear{{Balick}, {Frank}  \& {Liu}}{{Balick} et~al.}{2020}]{Balicketal2020}
{Balick} B.,  {Frank} A.,   {Liu} B.,  2020, \mn@doi [\apj] {10.3847/1538-4357/ab5651}, \href {https://ui.adsabs.harvard.edu/abs/2020ApJ...889...13B} {889, 13}

\bibitem[\protect\citeauthoryear{{Bear} \& {Soker}}{{Bear} \& {Soker}}{2025}]{BearSoker2025acc}
{Bear} E.,  {Soker} N.,  2025, \mn@doi [Research in Astronomy and Astrophysics] {10.1088/1674-4527/ada8ef}, \href {https://ui.adsabs.harvard.edu/abs/2025RAA....25b5010B} {25, 025010}

\bibitem[\protect\citeauthoryear{{Berm{\'u}dez-Bustamante} et~al.,}{{Berm{\'u}dez-Bustamante} et~al.}{2024}]{BermudezBustamanteetal2024}
{Berm{\'u}dez-Bustamante} L.~C.,  et~al., 2024, \mn@doi [\mnras] {10.1093/mnras/stae1841}, \href {https://ui.adsabs.harvard.edu/abs/2024MNRAS.533..464B} {533, 464}

\bibitem[\protect\citeauthoryear{{Bhattacharyya}, {Chamandy}, {Blackman}, {Frank}  \& {Liu}}{{Bhattacharyya} et~al.}{2025}]{Bhattacharyyaetal2025}
{Bhattacharyya} S.,  {Chamandy} L.,  {Blackman} E.~G.,  {Frank} A.,   {Liu} B.,  2025, arXiv e-prints, \href {https://ui.adsabs.harvard.edu/abs/2025arXiv250619547B} {p. arXiv:2506.19547}

\bibitem[\protect\citeauthoryear{{Blackman} \& {Lucchini}}{{Blackman} \& {Lucchini}}{2014}]{BlackmanLucchini2014}
{Blackman} E.~G.,  {Lucchini} S.,  2014, \mn@doi [\mnras] {10.1093/mnrasl/slu001}, \href {https://ui.adsabs.harvard.edu/abs/2014MNRAS.440L..16B} {440, L16}

\bibitem[\protect\citeauthoryear{{Bujarrabal}, {Castro-Carrizo}, {Alcolea}  \& {S{\'a}nchez Contreras}}{{Bujarrabal} et~al.}{2001}]{Bujarrabaletal2001}
{Bujarrabal} V.,  {Castro-Carrizo} A.,  {Alcolea} J.,   {S{\'a}nchez Contreras} C.,  2001, \mn@doi [\aap] {10.1051/0004-6361:20011090}, \href {https://ui.adsabs.harvard.edu/abs/2001A&A...377..868B} {377, 868}

\bibitem[\protect\citeauthoryear{{Chamandy} et~al.,}{{Chamandy} et~al.}{2018}]{Chamandyetal2018}
{Chamandy} L.,  et~al., 2018, \mn@doi [\mnras] {10.1093/mnras/sty1950}, \href {https://ui.adsabs.harvard.edu/abs/2018MNRAS.480.1898C} {480, 1898}

\bibitem[\protect\citeauthoryear{{Chamandy}, {Blackman}, {Frank}, {Carroll-Nellenback}, {Zou}  \& {Tu}}{{Chamandy} et~al.}{2019}]{Chamandyetal2019}
{Chamandy} L.,  {Blackman} E.~G.,  {Frank} A.,  {Carroll-Nellenback} J.,  {Zou} Y.,   {Tu} Y.,  2019, \mn@doi [\mnras] {10.1093/mnras/stz2813}, \href {https://ui.adsabs.harvard.edu/abs/2019MNRAS.490.3727C} {490, 3727}

\bibitem[\protect\citeauthoryear{{Chamandy}, {Carroll-Nellenback}, {Blackman}, {Frank}, {Tu}, {Liu}, {Zou}  \& {Nordhaus}}{{Chamandy} et~al.}{2024}]{Chamandyetal2024}
{Chamandy} L.,  {Carroll-Nellenback} J.,  {Blackman} E.~G.,  {Frank} A.,  {Tu} Y.,  {Liu} B.,  {Zou} Y.,   {Nordhaus} J.,  2024, \mn@doi [\mnras] {10.1093/mnras/stae036}, \href {https://ui.adsabs.harvard.edu/abs/2024MNRAS.528..234C} {528, 234}

\bibitem[\protect\citeauthoryear{{Chu}, {Jacoby}  \& {Arendt}}{{Chu} et~al.}{1987}]{Chuetal1987}
{Chu} Y.-H.,  {Jacoby} G.~H.,   {Arendt} R.,  1987, \mn@doi [\apjs] {10.1086/191207}, \href {https://ui.adsabs.harvard.edu/abs/1987ApJS...64..529C} {64, 529}

\bibitem[\protect\citeauthoryear{{Clairmont}, {Steffen}  \& {Koning}}{{Clairmont} et~al.}{2022}]{Clairmontetal2022}
{Clairmont} R.,  {Steffen} W.,   {Koning} N.,  2022, \mn@doi [\mnras] {10.1093/mnras/stac2375}, \href {https://ui.adsabs.harvard.edu/abs/2022MNRAS.516.2711C} {516, 2711}

\bibitem[\protect\citeauthoryear{{Corradi} \& {Schwarz}}{{Corradi} \& {Schwarz}}{1995}]{CorradiSchwarz1995}
{Corradi} R.~L.~M.,  {Schwarz} H.~E.,  1995, \aap, \href {https://ui.adsabs.harvard.edu/abs/1995A&A...293..871C} {293, 871}

\bibitem[\protect\citeauthoryear{{Danehkar}}{{Danehkar}}{2022}]{Danehkar2022}
{Danehkar} A.,  2022, \mn@doi [\apjs] {10.3847/1538-4365/ac5cca}, \href {https://ui.adsabs.harvard.edu/abs/2022ApJS..260...14D} {260, 14}

\bibitem[\protect\citeauthoryear{{Derlopa}, {Akras}, {Amram}, {Boumis}, {Chiotellis}  \& {de Oliveira}}{{Derlopa} et~al.}{2024}]{Derlopaetal2024}
{Derlopa} S.,  {Akras} S.,  {Amram} P.,  {Boumis} P.,  {Chiotellis} A.,   {de Oliveira} C.~M.,  2024, \mn@doi [\mnras] {10.1093/mnras/stae1013}, \href {https://ui.adsabs.harvard.edu/abs/2024MNRAS.530.3327D} {530, 3327}

\bibitem[\protect\citeauthoryear{{Estrella-Trujillo}, {Hern{\'a}ndez-Mart{\'\i}nez}, {Vel{\'a}zquez}, {Esquivel}  \& {Raga}}{{Estrella-Trujillo} et~al.}{2019}]{EstrellaTrujilloetal2019}
{Estrella-Trujillo} D.,  {Hern{\'a}ndez-Mart{\'\i}nez} L.,  {Vel{\'a}zquez} P.~F.,  {Esquivel} A.,   {Raga} A.~C.,  2019, \mn@doi [\apj] {10.3847/1538-4357/ab12e1}, \href {https://ui.adsabs.harvard.edu/abs/2019ApJ...876...29E} {876, 29}

\bibitem[\protect\citeauthoryear{{Gagnier} \& {Pejcha}}{{Gagnier} \& {Pejcha}}{2024}]{GagnierPejcha2024}
{Gagnier} D.,  {Pejcha} O.,  2024, \mn@doi [\aap] {10.1051/0004-6361/202348383}, \href {https://ui.adsabs.harvard.edu/abs/2024A&A...683A...4G} {683, A4}

\bibitem[\protect\citeauthoryear{{Garcia-Segura}}{{Garcia-Segura}}{1997}]{GarciaSegura1997}
{Garcia-Segura} G.,  1997, \mn@doi [\apjl] {10.1086/316796}, \href {https://ui.adsabs.harvard.edu/abs/1997ApJ...489L.189G} {489, L189}

\bibitem[\protect\citeauthoryear{{Garc{\'\i}a-Segura} \& {L{\'o}pez}}{{Garc{\'\i}a-Segura} \& {L{\'o}pez}}{2000}]{GarciaSeguraLopez2000}
{Garc{\'\i}a-Segura} G.,  {L{\'o}pez} J.~A.,  2000, \mn@doi [\apj] {10.1086/317186}, \href {https://ui.adsabs.harvard.edu/abs/2000ApJ...544..336G} {544, 336}

\bibitem[\protect\citeauthoryear{{Garc{\'\i}a-Segura}, {L{\'o}pez}  \& {Franco}}{{Garc{\'\i}a-Segura} et~al.}{2005}]{GarciaSeguraetal2005}
{Garc{\'\i}a-Segura} G.,  {L{\'o}pez} J.~A.,   {Franco} J.,  2005, \mn@doi [\apj] {10.1086/426110}, \href {https://ui.adsabs.harvard.edu/abs/2005ApJ...618..919G} {618, 919}

\bibitem[\protect\citeauthoryear{{Garc{\'\i}a-Segura}, {Taam}  \& {Ricker}}{{Garc{\'\i}a-Segura} et~al.}{2021}]{GarciaSeguraetal2021}
{Garc{\'\i}a-Segura} G.,  {Taam} R.~E.,   {Ricker} P.~M.,  2021, \mn@doi [\apj] {10.3847/1538-4357/abfc4e}, \href {https://ui.adsabs.harvard.edu/abs/2021ApJ...914..111G} {914, 111}

\bibitem[\protect\citeauthoryear{{Garc{\'\i}a-Segura}, {Taam}  \& {Ricker}}{{Garc{\'\i}a-Segura} et~al.}{2022}]{GarciaSeguraetal2022}
{Garc{\'\i}a-Segura} G.,  {Taam} R.~E.,   {Ricker} P.~M.,  2022, \mn@doi [\mnras] {10.1093/mnras/stac2824}, \href {https://ui.adsabs.harvard.edu/abs/2022MNRAS.517.3822G} {517, 3822}

\bibitem[\protect\citeauthoryear{{Glanz} \& {Perets}}{{Glanz} \& {Perets}}{2021a}]{GlanzPerets2021a}
{Glanz} H.,  {Perets} H.~B.,  2021a, \mn@doi [\mnras] {10.1093/mnras/staa3242}, \href {https://ui.adsabs.harvard.edu/abs/2021MNRAS.500.1921G} {500, 1921}

\bibitem[\protect\citeauthoryear{{Glanz} \& {Perets}}{{Glanz} \& {Perets}}{2021b}]{GlanzPerets2021b}
{Glanz} H.,  {Perets} H.~B.,  2021b, \mn@doi [\mnras] {10.1093/mnras/stab2291}, \href {https://ui.adsabs.harvard.edu/abs/2021MNRAS.507.2659G} {507, 2659}

\bibitem[\protect\citeauthoryear{{Gonz{\'a}lez-Bol{\'\i}var}, {De Marco}, {Lau}, {Hirai}  \& {Price}}{{Gonz{\'a}lez-Bol{\'\i}var} et~al.}{2022}]{GonzalezBolivaretal2022}
{Gonz{\'a}lez-Bol{\'\i}var} M.,  {De Marco} O.,  {Lau} M. Y.~M.,  {Hirai} R.,   {Price} D.~J.,  2022, \mn@doi [\mnras] {10.1093/mnras/stac2301}, \href {https://ui.adsabs.harvard.edu/abs/2022MNRAS.517.3181G} {517, 3181}

\bibitem[\protect\citeauthoryear{{Gonz{\'a}lez-Bol{\'\i}var}, {De Marco}, {Berm{\'u}dez-Bustamante}, {Siess}  \& {Price}}{{Gonz{\'a}lez-Bol{\'\i}var} et~al.}{2024}]{GonzalezBolivaretal2024}
{Gonz{\'a}lez-Bol{\'\i}var} M.,  {De Marco} O.,  {Berm{\'u}dez-Bustamante} L.~C.,  {Siess} L.,   {Price} D.~J.,  2024, \mn@doi [\mnras] {10.1093/mnras/stad3748}, \href {https://ui.adsabs.harvard.edu/abs/2024MNRAS.527.9145G} {527, 9145}

\bibitem[\protect\citeauthoryear{{Grichener}, {Cohen}  \& {Soker}}{{Grichener} et~al.}{2021}]{GrichenerCohenSoker2021}
{Grichener} A.,  {Cohen} C.,   {Soker} N.,  2021, \mn@doi [\apj] {10.3847/1538-4357/ac23dd}, \href {https://ui.adsabs.harvard.edu/abs/2021ApJ...922...61G} {922, 61}

\bibitem[\protect\citeauthoryear{{Guerrero}, {Suzett Rechy-Garc{\'\i}a}  \& {Ortiz}}{{Guerrero} et~al.}{2020}]{Guerreroetal2020}
{Guerrero} M.~A.,  {Suzett Rechy-Garc{\'\i}a} J.,   {Ortiz} R.,  2020, \mn@doi [\apj] {10.3847/1538-4357/ab61fa}, \href {https://ui.adsabs.harvard.edu/abs/2020ApJ...890...50G} {890, 50}

\bibitem[\protect\citeauthoryear{{Gurjar}, {Chamandy}, {Zou}, {Blackman}, {Liu}  \& {Nordhaus}}{{Gurjar} et~al.}{2024}]{Gurjareta2024eas}
{Gurjar} D.,  {Chamandy} L.~R.,  {Zou} A.,  {Blackman} E.,  {Liu} B.,   {Nordhaus} J.,  2024, in EAS2024, European Astronomical Society Annual Meeting. p.~438

\bibitem[\protect\citeauthoryear{{Hillel}, {Schreier}  \& {Soker}}{{Hillel} et~al.}{2022}]{Hilleletal2022}
{Hillel} S.,  {Schreier} R.,   {Soker} N.,  2022, \mn@doi [\mnras] {10.1093/mnras/stac1341}, \href {https://ui.adsabs.harvard.edu/abs/2022MNRAS.514.3212H} {514, 3212}

\bibitem[\protect\citeauthoryear{{Hillel}, {Schreier}  \& {Soker}}{{Hillel} et~al.}{2023}]{Hilleletal2023}
{Hillel} S.,  {Schreier} R.,   {Soker} N.,  2023, \mn@doi [\apj] {10.3847/1538-4357/acf19a}, \href {https://ui.adsabs.harvard.edu/abs/2023ApJ...955....7H} {955, 7}

\bibitem[\protect\citeauthoryear{{Hillwig}, {Frew}, {Reindl}, {Rotter}, {Webb}  \& {Margheim}}{{Hillwig} et~al.}{2017}]{Hillwigetal2017}
{Hillwig} T.~C.,  {Frew} D.~J.,  {Reindl} N.,  {Rotter} H.,  {Webb} A.,   {Margheim} S.,  2017, \mn@doi [\aj] {10.3847/1538-3881/153/1/24}, \href {https://ui.adsabs.harvard.edu/abs/2017AJ....153...24H} {153, 24}

\bibitem[\protect\citeauthoryear{{Iaconi}, {Reichardt}, {Staff}, {De Marco}, {Passy}, {Price}, {Wurster}  \& {Herwig}}{{Iaconi} et~al.}{2017}]{Iaconietal2017b}
{Iaconi} R.,  {Reichardt} T.,  {Staff} J.,  {De Marco} O.,  {Passy} J.-C.,  {Price} D.,  {Wurster} J.,   {Herwig} F.,  2017, \mn@doi [\mnras] {10.1093/mnras/stw2377}, \href {https://ui.adsabs.harvard.edu/abs/2017MNRAS.464.4028I} {464, 4028}

\bibitem[\protect\citeauthoryear{{Jermyn} et~al.,}{{Jermyn} et~al.}{2023}]{Jermyn2023}
{Jermyn} A.~S.,  et~al., 2023, \mn@doi [\apjs] {10.3847/1538-4365/acae8d}, \href {https://ui.adsabs.harvard.edu/abs/2023ApJS..265...15J} {265, 15}

\bibitem[\protect\citeauthoryear{{Jones}}{{Jones}}{2020}]{Jones2020Galax}
{Jones} D.,  2020, \mn@doi [Galaxies] {10.3390/galaxies8020028}, \href {https://ui.adsabs.harvard.edu/abs/2020Galax...8...28J} {8, 28}

\bibitem[\protect\citeauthoryear{{Jones}}{{Jones}}{2024}]{Jones2025}
{Jones} D.,  2024, \mn@doi [arXiv e-prints] {10.48550/arXiv.2411.06831}, \href {https://ui.adsabs.harvard.edu/abs/2024arXiv241106831J} {p. arXiv:2411.06831}

\bibitem[\protect\citeauthoryear{{Jones}, {Boffin}, {Rodr{\'\i}guez-Gil}, {Wesson}, {Corradi}, {Miszalski}  \& {Mohamed}}{{Jones} et~al.}{2015}]{Jonesetal2015}
{Jones} D.,  {Boffin} H.~M.~J.,  {Rodr{\'\i}guez-Gil} P.,  {Wesson} R.,  {Corradi} R.~L.~M.,  {Miszalski} B.,   {Mohamed} S.,  2015, \mn@doi [\aap] {10.1051/0004-6361/201425454}, \href {https://ui.adsabs.harvard.edu/abs/2015A&A...580A..19J} {580, A19}

\bibitem[\protect\citeauthoryear{{Jones} et~al.,}{{Jones} et~al.}{2020}]{Jonesetal2020}
{Jones} D.,  et~al., 2020, \mn@doi [\aap] {10.1051/0004-6361/202038778}, \href {https://ui.adsabs.harvard.edu/abs/2020A&A...642A.108J} {642, A108}

\bibitem[\protect\citeauthoryear{{Jones} et~al.,}{{Jones} et~al.}{2022}]{Jonesetal2022}
{Jones} D.,  et~al., 2022, \mn@doi [\mnras] {10.1093/mnras/stab3736}, \href {https://ui.adsabs.harvard.edu/abs/2022MNRAS.510.3102J} {510, 3102}

\bibitem[\protect\citeauthoryear{{Jones}, {Hillwig}  \& {Reindl}}{{Jones} et~al.}{2023}]{Jonesetal2023}
{Jones} D.,  {Hillwig} T.~C.,   {Reindl} N.,  2023, in {Manteiga} M.,  {Bellot} L.,  {Benavidez} P.,  {de Lorenzo-C{\'a}ceres} A.,  {Fuente} M.~A.,  {Mart{\'\i}nez} M.~J.,  {V{\'a}zquez Acosta} M.,   {Dafonte} C.,  eds, Highlights on Spanish Astrophysics XI. p.~216 (\mn@eprint {arXiv} {2304.06355}), \mn@doi{10.48550/arXiv.2304.06355}

\bibitem[\protect\citeauthoryear{{Kashi}, {Michaelis}  \& {Kaminetsky}}{{Kashi} et~al.}{2022}]{Kashietal2022}
{Kashi} A.,  {Michaelis} A.,   {Kaminetsky} Y.,  2022, \mn@doi [\mnras] {10.1093/mnras/stac1912}, \href {https://ui.adsabs.harvard.edu/abs/2022MNRAS.516.3193K} {516, 3193}

\bibitem[\protect\citeauthoryear{{Kimeswenger} et~al.,}{{Kimeswenger} et~al.}{2021}]{Kimeswengeretal2021}
{Kimeswenger} S.,  et~al., 2021, \mn@doi [\aap] {10.1051/0004-6361/202039787}, \href {https://ui.adsabs.harvard.edu/abs/2021A&A...656A.145K} {656, A145}

\bibitem[\protect\citeauthoryear{{Kuruwita}, {Staff}  \& {De Marco}}{{Kuruwita} et~al.}{2016}]{Kuruwitaetal2016}
{Kuruwita} R.~L.,  {Staff} J.,   {De Marco} O.,  2016, \mn@doi [\mnras] {10.1093/mnras/stw1414}, \href {https://ui.adsabs.harvard.edu/abs/2016MNRAS.461..486K} {461, 486}

\bibitem[\protect\citeauthoryear{{Landri}, {Ricker}, {Renzo}, {Rau}  \& {Vigna-G{\'o}mez}}{{Landri} et~al.}{2024}]{Landrietal2024}
{Landri} C.,  {Ricker} P.~M.,  {Renzo} M.,  {Rau} S.,   {Vigna-G{\'o}mez} A.,  2024, \mn@doi [arXiv e-prints] {10.48550/arXiv.2407.15932}, \href {https://ui.adsabs.harvard.edu/abs/2024arXiv240715932L} {p. arXiv:2407.15932}

\bibitem[\protect\citeauthoryear{{Lau}, {Hirai}, {Gonz{\'a}lez-Bol{\'\i}var}, {Price}, {De Marco}  \& {Mandel}}{{Lau} et~al.}{2022a}]{Lauetal2022a}
{Lau} M. Y.~M.,  {Hirai} R.,  {Gonz{\'a}lez-Bol{\'\i}var} M.,  {Price} D.~J.,  {De Marco} O.,   {Mandel} I.,  2022a, \mn@doi [\mnras] {10.1093/mnras/stac049}, \href {https://ui.adsabs.harvard.edu/abs/2022MNRAS.512.5462L} {512, 5462}

\bibitem[\protect\citeauthoryear{{Lau}, {Hirai}, {Price}  \& {Mandel}}{{Lau} et~al.}{2022b}]{Lauetal2022b}
{Lau} M. Y.~M.,  {Hirai} R.,  {Price} D.~J.,   {Mandel} I.,  2022b, \mn@doi [\mnras] {10.1093/mnras/stac2490}, \href {https://ui.adsabs.harvard.edu/abs/2022MNRAS.516.4669L} {516, 4669}

\bibitem[\protect\citeauthoryear{{Law-Smith} et~al.,}{{Law-Smith} et~al.}{2020}]{LawSmithetal2020}
{Law-Smith} J. A.~P.,  et~al., 2020, \mn@doi [arXiv e-prints] {10.48550/arXiv.2011.06630}, \href {https://ui.adsabs.harvard.edu/abs/2020arXiv201106630L} {p. arXiv:2011.06630}

\bibitem[\protect\citeauthoryear{{Livio}, {Soker}, {de Kool}  \& {Savonije}}{{Livio} et~al.}{1986}]{Livio1986}
{Livio} M.,  {Soker} N.,  {de Kool} M.,   {Savonije} G.~J.,  1986, \mn@doi [\mnras] {10.1093/mnras/222.2.235}, \href {https://ui.adsabs.harvard.edu/abs/1986MNRAS.222..235L} {222, 235}

\bibitem[\protect\citeauthoryear{{L{\'o}pez-C{\'a}mara}, {De Colle}  \& {Moreno M{\'e}ndez}}{{L{\'o}pez-C{\'a}mara} et~al.}{2019}]{LopezCamaraetal2019}
{L{\'o}pez-C{\'a}mara} D.,  {De Colle} F.,   {Moreno M{\'e}ndez} E.,  2019, \mn@doi [\mnras] {10.1093/mnras/sty2959}, \href {https://ui.adsabs.harvard.edu/abs/2019MNRAS.482.3646L} {482, 3646}

\bibitem[\protect\citeauthoryear{{L{\'o}pez-C{\'a}mara}, {Moreno M{\'e}ndez}  \& {De Colle}}{{L{\'o}pez-C{\'a}mara} et~al.}{2020}]{LopezCamaraetal2020MN}
{L{\'o}pez-C{\'a}mara} D.,  {Moreno M{\'e}ndez} E.,   {De Colle} F.,  2020, \mn@doi [\mnras] {10.1093/mnras/staa1983}, \href {https://ui.adsabs.harvard.edu/abs/2020MNRAS.497.2057L} {497, 2057}

\bibitem[\protect\citeauthoryear{{L{\'o}pez-C{\'a}mara}, {De Colle}, {Moreno M{\'e}ndez}, {Shiber}  \& {Iaconi}}{{L{\'o}pez-C{\'a}mara} et~al.}{2022}]{LopezCamaraetal2022}
{L{\'o}pez-C{\'a}mara} D.,  {De Colle} F.,  {Moreno M{\'e}ndez} E.,  {Shiber} S.,   {Iaconi} R.,  2022, \mn@doi [\mnras] {10.1093/mnras/stac932}, \href {https://ui.adsabs.harvard.edu/abs/2022MNRAS.513.3634L} {513, 3634}

\bibitem[\protect\citeauthoryear{{MacLeod} \& {Ramirez-Ruiz}}{{MacLeod} \& {Ramirez-Ruiz}}{2015}]{MacLeodRamirezRuiz2015}
{MacLeod} M.,  {Ramirez-Ruiz} E.,  2015, \mn@doi [\apjl] {10.1088/2041-8205/798/1/L19}, \href {https://ui.adsabs.harvard.edu/abs/2015ApJ...798L..19M} {798, L19}

\bibitem[\protect\citeauthoryear{{MacLeod}, {Antoni}, {Murguia-Berthier}, {Macias}  \& {Ramirez-Ruiz}}{{MacLeod} et~al.}{2017}]{MacLeodetal2017}
{MacLeod} M.,  {Antoni} A.,  {Murguia-Berthier} A.,  {Macias} P.,   {Ramirez-Ruiz} E.,  2017, \mn@doi [\apj] {10.3847/1538-4357/aa6117}, \href {https://ui.adsabs.harvard.edu/abs/2017ApJ...838...56M} {838, 56}

\bibitem[\protect\citeauthoryear{{Miranda}, {V{\'a}zquez}, {Olgu{\'\i}n}, {Guill{\'e}n}  \& {Mat{\'\i}as}}{{Miranda} et~al.}{2024}]{Mirandaetal2024}
{Miranda} L.~F.,  {V{\'a}zquez} R.,  {Olgu{\'\i}n} L.,  {Guill{\'e}n} P.~F.,   {Mat{\'\i}as} J.~M.,  2024, \mn@doi [\aap] {10.1051/0004-6361/202348173}, \href {https://ui.adsabs.harvard.edu/abs/2024A&A...687A.123M} {687, A123}

\bibitem[\protect\citeauthoryear{{Miszalski}, {Manick}, {Van Winckel}  \& {Miko{\l}ajewska}}{{Miszalski} et~al.}{2019}]{Miszalski2019ic}
{Miszalski} B.,  {Manick} R.,  {Van Winckel} H.,   {Miko{\l}ajewska} J.,  2019, \mn@doi [\mnras] {10.1093/mnras/stz1315}, \href {https://ui.adsabs.harvard.edu/abs/2019MNRAS.487.1040M} {487, 1040}

\bibitem[\protect\citeauthoryear{{Moraga Baez}, {Kastner}, {Balick}, {Montez}  \& {Bublitz}}{{Moraga Baez} et~al.}{2023}]{MoragaBaezetal2023}
{Moraga Baez} P.,  {Kastner} J.~H.,  {Balick} B.,  {Montez} R.,   {Bublitz} J.,  2023, \mn@doi [\apj] {10.3847/1538-4357/aca401}, \href {https://ui.adsabs.harvard.edu/abs/2023ApJ...942...15M} {942, 15}

\bibitem[\protect\citeauthoryear{{Moreno M{\'e}ndez}, {L{\'o}pez-C{\'a}mara}  \& {De Colle}}{{Moreno M{\'e}ndez} et~al.}{2017}]{MorenoMendezetal2017}
{Moreno M{\'e}ndez} E.,  {L{\'o}pez-C{\'a}mara} D.,   {De Colle} F.,  2017, \mn@doi [\mnras] {10.1093/mnras/stx1385}, \href {https://ui.adsabs.harvard.edu/abs/2017MNRAS.470.2929M} {470, 2929}

\bibitem[\protect\citeauthoryear{{Morris}}{{Morris}}{1987}]{Morris1987}
{Morris} M.,  1987, \mn@doi [\pasp] {10.1086/132089}, \href {https://ui.adsabs.harvard.edu/abs/1987PASP...99.1115M} {99, 1115}

\bibitem[\protect\citeauthoryear{{Ohlmann}, {R{\"o}pke}, {Pakmor}  \& {Springel}}{{Ohlmann} et~al.}{2016}]{Ohlmannetal2016a}
{Ohlmann} S.~T.,  {R{\"o}pke} F.~K.,  {Pakmor} R.,   {Springel} V.,  2016, \mn@doi [\apjl] {10.3847/2041-8205/816/1/L9}, \href {https://ui.adsabs.harvard.edu/abs/2016ApJ...816L...9O} {816, L9}

\bibitem[\protect\citeauthoryear{{Orosz} et~al.,}{{Orosz} et~al.}{2017}]{Oroszetal2019}
{Orosz} G.,  et~al., 2017, \mn@doi [\aj] {10.3847/1538-3881/153/3/119}, \href {https://ui.adsabs.harvard.edu/abs/2017AJ....153..119O} {153, 119}

\bibitem[\protect\citeauthoryear{{Parker}}{{Parker}}{2022}]{Parker2022}
{Parker} Q.~A.,  2022, \mn@doi [Frontiers in Astronomy and Space Sciences] {10.3389/fspas.2022.895287}, \href {https://ui.adsabs.harvard.edu/abs/2022FrASS...9.5287P} {9, 895287}

\bibitem[\protect\citeauthoryear{{Parker}, {Boji{\v{c}}i{\'c}}  \& {Frew}}{{Parker} et~al.}{2016}]{Parkeretal2016}
{Parker} Q.~A.,  {Boji{\v{c}}i{\'c}} I.~S.,   {Frew} D.~J.,  2016, in Journal of Physics Conference Series. IOP, p. 032008 (\mn@eprint {arXiv} {1603.07042}), \mn@doi{10.1088/1742-6596/728/3/032008}

\bibitem[\protect\citeauthoryear{{Paxton}, {Bildsten}, {Dotter}, {Herwig}, {Lesaffre}  \& {Timmes}}{{Paxton} et~al.}{2011}]{Paxton2011}
{Paxton} B.,  {Bildsten} L.,  {Dotter} A.,  {Herwig} F.,  {Lesaffre} P.,   {Timmes} F.,  2011, \mn@doi [\apjs] {10.1088/0067-0049/192/1/3}, \href {https://ui.adsabs.harvard.edu/abs/2011ApJS..192....3P} {192, 3}

\bibitem[\protect\citeauthoryear{{Paxton} et~al.,}{{Paxton} et~al.}{2013}]{Paxton2013}
{Paxton} B.,  et~al., 2013, \mn@doi [\apjs] {10.1088/0067-0049/208/1/4}, \href {https://ui.adsabs.harvard.edu/abs/2013ApJS..208....4P} {208, 4}

\bibitem[\protect\citeauthoryear{{Paxton} et~al.,}{{Paxton} et~al.}{2015}]{Paxton2015}
{Paxton} B.,  et~al., 2015, \mn@doi [\apjs] {10.1088/0067-0049/220/1/15}, \href {https://ui.adsabs.harvard.edu/abs/2015ApJS..220...15P} {220, 15}

\bibitem[\protect\citeauthoryear{{Paxton} et~al.,}{{Paxton} et~al.}{2018}]{Paxton2018}
{Paxton} B.,  et~al., 2018, \mn@doi [\apjs] {10.3847/1538-4365/aaa5a8}, \href {https://ui.adsabs.harvard.edu/abs/2018ApJS..234...34P} {234, 34}

\bibitem[\protect\citeauthoryear{{Paxton} et~al.,}{{Paxton} et~al.}{2019}]{Paxton2019}
{Paxton} B.,  et~al., 2019, \mn@doi [\apjs] {10.3847/1538-4365/ab2241}, \href {https://ui.adsabs.harvard.edu/abs/2019ApJS..243...10P} {243, 10}

\bibitem[\protect\citeauthoryear{{Qi}, {Liu}  \& {Wang}}{{Qi} et~al.}{2023}]{Qietal2023}
{Qi} W.-Z.,  {Liu} D.-D.,   {Wang} B.,  2023, \mn@doi [Research in Astronomy and Astrophysics] {10.1088/1674-4527/aca235}, \href {https://ui.adsabs.harvard.edu/abs/2023RAA....23a5008Q} {23, 015008}

\bibitem[\protect\citeauthoryear{{Rechy-Garc{\'\i}a}, {Guerrero}, {Duarte Puertas}, {Chu}, {Toal{\'a}}  \& {Miranda}}{{Rechy-Garc{\'\i}a} et~al.}{2020}]{RechyGarciaetal2020}
{Rechy-Garc{\'\i}a} J.~S.,  {Guerrero} M.~A.,  {Duarte Puertas} S.,  {Chu} Y.~H.,  {Toal{\'a}} J.~A.,   {Miranda} L.~F.,  2020, \mn@doi [\mnras] {10.1093/mnras/stz3326}, \href {https://ui.adsabs.harvard.edu/abs/2020MNRAS.492.1957R} {492, 1957}

\bibitem[\protect\citeauthoryear{Ricker \& Taam}{Ricker \& Taam}{2007}]{Ricker2008}
Ricker P.~M.,  Taam R.~E.,  2007, \mn@doi [The Astrophysical Journal] {10.1086/526343}, 672, L41

\bibitem[\protect\citeauthoryear{{Rosselli-Calderon}, {Yarza}, {Murguia-Berthier}, {Rohoza}, {Everson}, {Antoni}, {MacLeod}  \& {Ramirez-Ruiz}}{{Rosselli-Calderon} et~al.}{2024}]{RosselliCalderon2024}
{Rosselli-Calderon} A.,  {Yarza} R.,  {Murguia-Berthier} A.,  {Rohoza} V.,  {Everson} R.~W.,  {Antoni} A.,  {MacLeod} M.,   {Ramirez-Ruiz} E.,  2024, \mn@doi [arXiv e-prints] {10.48550/arXiv.2404.08037}, \href {https://ui.adsabs.harvard.edu/abs/2024arXiv240408037R} {p. arXiv:2404.08037}

\bibitem[\protect\citeauthoryear{{Sahai} \& {Trauger}}{{Sahai} \& {Trauger}}{1998}]{SahaiTrauger1998}
{Sahai} R.,  {Trauger} J.~T.,  1998, \mn@doi [\aj] {10.1086/300504}, \href {https://ui.adsabs.harvard.edu/abs/1998AJ....116.1357S} {116, 1357}

\bibitem[\protect\citeauthoryear{{Sahai}, {Morris}, {S{\'a}nchez Contreras}  \& {Claussen}}{{Sahai} et~al.}{2007}]{Sahaietal2007}
{Sahai} R.,  {Morris} M.,  {S{\'a}nchez Contreras} C.,   {Claussen} M.,  2007, \mn@doi [\aj] {10.1086/522944}, \href {https://ui.adsabs.harvard.edu/abs/2007AJ....134.2200S} {134, 2200}

\bibitem[\protect\citeauthoryear{{Sahai}, {Claussen}, {S{\'a}nchez Contreras}, {Morris}  \& {Sarkar}}{{Sahai} et~al.}{2008}]{Sahaietal2008}
{Sahai} R.,  {Claussen} M.,  {S{\'a}nchez Contreras} C.,  {Morris} M.,   {Sarkar} G.,  2008, \mn@doi [\apj] {10.1086/587638}, \href {https://ui.adsabs.harvard.edu/abs/2008ApJ...680..483S} {680, 483}

\bibitem[\protect\citeauthoryear{{Sahai}, {Morris}  \& {Villar}}{{Sahai} et~al.}{2011}]{Sahaietal2011}
{Sahai} R.,  {Morris} M.~R.,   {Villar} G.~G.,  2011, \mn@doi [\aj] {10.1088/0004-6256/141/4/134}, \href {https://ui.adsabs.harvard.edu/abs/2011AJ....141..134S} {141, 134}

\bibitem[\protect\citeauthoryear{{Sahai}, {Vlemmings}  \& {Nyman}}{{Sahai} et~al.}{2017}]{Sahaietal2017}
{Sahai} R.,  {Vlemmings} W.~H.~T.,   {Nyman} L.~{\r{A}}.,  2017, \mn@doi [\apj] {10.3847/1538-4357/aa6d86}, \href {https://ui.adsabs.harvard.edu/abs/2017ApJ...841..110S} {841, 110}

\bibitem[\protect\citeauthoryear{{Sahai} et~al.,}{{Sahai} et~al.}{2024}]{Sahaietal2024}
{Sahai} R.,  et~al., 2024, arXiv e-prints, \href {https://ui.adsabs.harvard.edu/abs/2024arXiv240906038S} {p. arXiv:2409.06038}

\bibitem[\protect\citeauthoryear{{Schreier}, {Hillel}  \& {Soker}}{{Schreier} et~al.}{2023}]{Schreieretal2023}
{Schreier} R.,  {Hillel} S.,   {Soker} N.,  2023, \mn@doi [\mnras] {10.1093/mnras/stad360}, \href {https://ui.adsabs.harvard.edu/abs/2023MNRAS.520.4182S} {520, 4182}

\bibitem[\protect\citeauthoryear{{Schreier}, {Hillel}  \& {Soker}}{{Schreier} et~al.}{2025}]{Schreieretal2025}
{Schreier} R.,  {Hillel} S.,   {Soker} N.,  2025, \mn@doi [arXiv e-prints] {10.48550/arXiv.2501.09663}, \href {https://ui.adsabs.harvard.edu/abs/2025arXiv250109663S} {p. arXiv:2501.09663}

\bibitem[\protect\citeauthoryear{{Schwarz}, {Corradi}  \& {Melnick}}{{Schwarz} et~al.}{1992}]{Schwarzetal1992}
{Schwarz} H.~E.,  {Corradi} R.~L.~M.,   {Melnick} J.,  1992, \aaps, \href {https://ui.adsabs.harvard.edu/abs/1992A&AS...96...23S} {96, 23}

\bibitem[\protect\citeauthoryear{{Scolnic}, {Bear}  \& {Soker}}{{Scolnic} et~al.}{2025}]{Scolnicetal2025}
{Scolnic} A.,  {Bear} E.,   {Soker} N.,  2025, \mn@doi [\pasp] {10.1088/1538-3873/adb5c2}, \href {https://ui.adsabs.harvard.edu/abs/2025PASP..137c4201S} {137, 034201}

\bibitem[\protect\citeauthoryear{{Shiber} \& {Iaconi}}{{Shiber} \& {Iaconi}}{2024}]{ShiberIaconi2024}
{Shiber} S.,  {Iaconi} R.,  2024, \mn@doi [\mnras] {10.1093/mnras/stae1500}, \href {https://ui.adsabs.harvard.edu/abs/2024MNRAS.532..692S} {532, 692}

\bibitem[\protect\citeauthoryear{{Shiber} \& {Soker}}{{Shiber} \& {Soker}}{2018}]{ShiberSoker2018}
{Shiber} S.,  {Soker} N.,  2018, \mn@doi [\mnras] {10.1093/mnras/sty843}, \href {https://ui.adsabs.harvard.edu/abs/2018MNRAS.477.2584S} {477, 2584}

\bibitem[\protect\citeauthoryear{{Shiber}, {Schreier}  \& {Soker}}{{Shiber} et~al.}{2016}]{Shiberetal2016}
{Shiber} S.,  {Schreier} R.,   {Soker} N.,  2016, \mn@doi [Research in Astronomy and Astrophysics] {10.1088/1674-4527/16/7/117}, \href {https://ui.adsabs.harvard.edu/abs/2016RAA....16..117S} {16, 117}

\bibitem[\protect\citeauthoryear{{Shiber}, {Iaconi}, {De Marco}  \& {Soker}}{{Shiber} et~al.}{2019}]{Shiberetal2019}
{Shiber} S.,  {Iaconi} R.,  {De Marco} O.,   {Soker} N.,  2019, \mn@doi [\mnras] {10.1093/mnras/stz2013}, \href {https://ui.adsabs.harvard.edu/abs/2019MNRAS.488.5615S} {488, 5615}

\bibitem[\protect\citeauthoryear{{Soker}}{{Soker}}{1990}]{Soker1990AJ}
{Soker} N.,  1990, \mn@doi [\aj] {10.1086/115465}, \href {https://ui.adsabs.harvard.edu/abs/1990AJ.....99.1869S} {99, 1869}

\bibitem[\protect\citeauthoryear{{Soker}}{{Soker}}{2016}]{Soker2016Rev}
{Soker} N.,  2016, \mn@doi [\nar] {10.1016/j.newar.2016.08.002}, \href {https://ui.adsabs.harvard.edu/abs/2016NewAR..75....1S} {75, 1}

\bibitem[\protect\citeauthoryear{{Soker}}{{Soker}}{2020}]{Soker2020Galax}
{Soker} N.,  2020, \mn@doi [Galaxies] {10.3390/galaxies8010026}, \href {https://ui.adsabs.harvard.edu/abs/2020Galax...8...26S} {8, 26}

\bibitem[\protect\citeauthoryear{{Soker}}{{Soker}}{2022}]{Soker2022Rev}
{Soker} N.,  2022, \mn@doi [Research in Astronomy and Astrophysics] {10.1088/1674-4527/ac9782}, \href {https://ui.adsabs.harvard.edu/abs/2022RAA....22l2003S} {22, 122003}

\bibitem[\protect\citeauthoryear{{Soker}}{{Soker}}{2023}]{Soker2023RAA}
{Soker} N.,  2023, \mn@doi [Research in Astronomy and Astrophysics] {10.1088/1674-4527/acdfa8}, \href {https://ui.adsabs.harvard.edu/abs/2023RAA....23i5002S} {23, 095002}

\bibitem[\protect\citeauthoryear{{Soker}}{{Soker}}{2025a}]{Soker2025WDfeed}
{Soker} N.,  2025a, arXiv e-prints, \href {https://ui.adsabs.harvard.edu/abs/2025arXiv250522621S} {p. arXiv:2505.22621}

\bibitem[\protect\citeauthoryear{{Soker}}{{Soker}}{2025b}]{Soker2025Robust}
{Soker} N.,  2025b, \mn@doi [Research in Astronomy and Astrophysics] {10.1088/1674-4527/adb15b}, \href {https://ui.adsabs.harvard.edu/abs/2025RAA....25b5023S} {25, 025023}

\bibitem[\protect\citeauthoryear{{Staff}, {De Marco}, {Macdonald}, {Galaviz}, {Passy}, {Iaconi}  \& {Low}}{{Staff} et~al.}{2016}]{Staffetal2016MN}
{Staff} J.~E.,  {De Marco} O.,  {Macdonald} D.,  {Galaviz} P.,  {Passy} J.-C.,  {Iaconi} R.,   {Low} M.-M.~M.,  2016, \mn@doi [\mnras] {10.1093/mnras/stv2548}, \href {https://ui.adsabs.harvard.edu/abs/2016MNRAS.455.3511S} {455, 3511}

\bibitem[\protect\citeauthoryear{{Tafoya}, {Orosz}, {Vlemmings}, {Sahai}  \& {P{\'e}rez-S{\'a}nchez}}{{Tafoya} et~al.}{2019}]{Tafoyaetal2019}
{Tafoya} D.,  {Orosz} G.,  {Vlemmings} W.~H.~T.,  {Sahai} R.,   {P{\'e}rez-S{\'a}nchez} A.~F.,  2019, \mn@doi [\aap] {10.1051/0004-6361/201834632}, \href {https://ui.adsabs.harvard.edu/abs/2019A&A...629A...8T} {629, A8}

\bibitem[\protect\citeauthoryear{{Tocknell}, {De Marco}  \& {Wardle}}{{Tocknell} et~al.}{2014}]{Tocknelletal2014}
{Tocknell} J.,  {De Marco} O.,   {Wardle} M.,  2014, \mn@doi [\mnras] {10.1093/mnras/stu079}, \href {https://ui.adsabs.harvard.edu/abs/2014MNRAS.439.2014T} {439, 2014}

\bibitem[\protect\citeauthoryear{{Vetter}, {Roepke}, {Schneider}, {Pakmor}, {Ohlmann}, {Lau}  \& {Andrassy}}{{Vetter} et~al.}{2024}]{Vetteretal2024}
{Vetter} M.,  {Roepke} F.~K.,  {Schneider} F.~R.~N.,  {Pakmor} R.,  {Ohlmann} S.~T.,  {Lau} M.~Y.~M.,   {Andrassy} R.,  2024, \mn@doi [arXiv e-prints] {10.48550/arXiv.2410.07841}, \href {https://ui.adsabs.harvard.edu/abs/2024arXiv241007841V} {p. arXiv:2410.07841}

\bibitem[\protect\citeauthoryear{{Vetter} et~al.,}{{Vetter} et~al.}{2025}]{Vetteretal2025}
{Vetter} M.,  et~al., 2025, \mn@doi [arXiv e-prints] {10.48550/arXiv.2504.12213}, \href {https://ui.adsabs.harvard.edu/abs/2025arXiv250412213V} {p. arXiv:2504.12213}

\bibitem[\protect\citeauthoryear{{Xue}, {Qin}, {Yuan}, {Guo}, {Li}, {Zhang}, {Wang}  \& {Wu}}{{Xue} et~al.}{2025}]{Xueetal2025RAA}
{Xue} Y.-W.,  {Qin} Y.,  {Yuan} L.,  {Guo} W.-H.,  {Li} J.-Q.,  {Zhang} Y.-Q.,  {Wang} Z.-Y.,   {Wu} D.-H.,  2025, \mn@doi [Research in Astronomy and Astrophysics] {10.1088/1674-4527/adc64f}, \href {https://ui.adsabs.harvard.edu/abs/2025RAA....25d5009X} {25, 045009}

\bibitem[\protect\citeauthoryear{{Zheng}, {Li}, {Wei}  \& {Jin}}{{Zheng} et~al.}{2025}]{Zhengetal2025RAA}
{Zheng} T.-C.,  {Li} X.-D.,  {Wei} D.-M.,   {Jin} Z.-P.,  2025, \mn@doi [Research in Astronomy and Astrophysics] {10.1088/1674-4527/adcf83}, \href {https://ui.adsabs.harvard.edu/abs/2025RAA....25f5001Z} {25, 065001}

\bibitem[\protect\citeauthoryear{{Zhou}, {Wang}, {Meng}, {Liu}  \& {Zhao}}{{Zhou} et~al.}{2015}]{Zhouetal2015RAA}
{Zhou} W.-H.,  {Wang} B.,  {Meng} X.-C.,  {Liu} D.-D.,   {Zhao} G.,  2015, \mn@doi [Research in Astronomy and Astrophysics] {10.1088/1674-4527/15/10/007}, \href {https://ui.adsabs.harvard.edu/abs/2015RAA....15.1701Z} {15, 1701}

\bibitem[\protect\citeauthoryear{{Zhu}, {L{\"u}}, {Lu}  \& {He}}{{Zhu} et~al.}{2023}]{Zhuetal2023}
{Zhu} C.-H.,  {L{\"u}} G.-L.,  {Lu} X.-Z.,   {He} J.,  2023, \mn@doi [Research in Astronomy and Astrophysics] {10.1088/1674-4527/acafc7}, \href {https://ui.adsabs.harvard.edu/abs/2023RAA....23b5021Z} {23, 025021}

\bibitem[\protect\citeauthoryear{{Zou}, {Chamandy}, {Carroll-Nellenback}, {Blackman}  \& {Frank}}{{Zou} et~al.}{2022}]{Zouetal2022}
{Zou} Y.,  {Chamandy} L.,  {Carroll-Nellenback} J.,  {Blackman} E.~G.,   {Frank} A.,  2022, \mn@doi [\mnras] {10.1093/mnras/stac1529}, \href {https://ui.adsabs.harvard.edu/abs/2022MNRAS.514.3041Z} {514, 3041}

\makeatother
\end{thebibliography}

\end{document}